\documentclass[fleqn,usenatbib]{mnras}
\usepackage[T1]{fontenc}
\usepackage{ae,aecompl}
\usepackage{graphicx}
\usepackage{amsmath}
\usepackage{supertabular}
\usepackage{booktabs}
\usepackage[normalem]{ulem}
\usepackage{xspace}
\usepackage{dcolumn}
 \newcolumntype{.}{D{.}{.}{-1}}
 \newcolumntype{p}{D{!}{\pm}{-1}}
 \newcolumntype{P}{D{.}{.}{5}}

\newcommand{\arcdeg}[0]{$^\circ$\xspace}
\newcommand{\kms}[0]{km~s$^{-1}$\xspace}
\newcommand{\sersic}[0]{S\'{e}rsic}
\newcommand{\targets}[0]{178}
\newcommand{\etal}[0]{et\ al.\xspace}
\newcommand{\diskfit}[0]{\textsc{diskfit}\xspace}
\newcommand{\galfit}[0]{\textsc{galfit3}\xspace}
\newcommand{\ha}[0]{H$\alpha$}
\newcommand{\hb}[0]{H$\beta$}
\newcommand{\MXL}{$\Upsilon_\times$\xspace}
\title[Rotationally Supported SAMI Survey Galaxies]{The SAMI Galaxy Survey: 
Gas Streaming and Dynamical M/L in Rotationally Supported Systems}
\author[G. Cecil et al.]{G.\ Cecil$^{1,2}$\thanks{Email: cecil@unc.edu (GC)},
L.\ M.\ R.\ Fogarty$^{2,3}$,
S.\ Richards$^{2,3,4}$,
J.\ Bland-Hawthorn$^{2}$,
R.\ Lange$^{5}$,
\newauthor
A.\ Moffett$^{5}$,
B.\ Catinella$^{5,6}$,
L.\ Cortese$^{5,6}$,
I.-T.\ Ho$^{7}$,
E.\ N.\ Taylor$^{8}$,
J.\ J.\ Bryant$^{2,3,4}$,
\newauthor
J.\ T.\ Allen$^{2,3}$,
S. M. Sweet$^{9}$,
S.\ M.\ Croom$^{2,3}$,
S.\ P.\ Driver$^{5}$,
M.\ Goodwin$^{4}$,
L.\ Kelvin$^{10}$,
\newauthor
A.\ W.\ Green$^{4}$,
I.\ S.\ Konstantopoulos$^{11}$,
M.\ S.\ Owers$^{4,12}$,
J.\ S.\ Lawrence$^{4}$,
\newauthor
N.\ P.\ F.\ Lorente$^{4}$
\\
$^{1}$ Dept.\ of Physics and Astronomy, University of North Carolina at Chapel Hill, NC 27510 \\
$^{2}$ Sydney Institute for Astronomy, School of Physics, University of Sydney, NSW 2006, Australia \\ 
$^{3}$ ARC Centre of Excellence for All-Sky Astrophysics \\
$^{4}$ Australian Astronomical Observatory, PO Box 915, North Ryde, NSW 1670, Australia \\
$^{5}$ International Centre for Radio Astronomy Research (ICRAR), School of Physics, The University of Western Australia, Perth, WA 6009,\\
Australia \\
$^{6}$ Centre for Astrophysics and Supercomputing, Swinburne University of Technology, Melbourne, VIC 3122, Australia \\
$^{7}$ Institute for Astronomy, University of Hawaii, 2680 Woodlawn Drive, Honolulu, HI 96822, USA \\
$^{8}$ School of Physics, The University of Melbourne, Parkville, VIC 3010, Australia \\
$^{9}$ Research School of Astronomy and Astrophysics, The Australian National University, Canberra, ACT 2611, Australia \\
$^{10}$ Institute for Astro- and Particle-Physics, University of Innsbruck, Innsbruck, Austria \\
$^{11}$ Envizi Group Suite 213, National Innovation Centre, Australian Technology Park, 4 Cornwallis Street, Eveleigh NSW 2015, Australia \\
$^{12}$ Dept.\ of Physics and Astronomy, Macquarie University, NSW 2109, Australia 
}
\date{Accepted 2015 November 08. Received 2015 October 14; in original form 2015 July 09}
\pubyear{2015}

\begin{document}
\label{firstpage}
\pagerange{\pageref{firstpage}--\pageref{lastpage}}
\maketitle

\begin{abstract}
Line-of-sight velocities of gas and stars can constrain dark matter (DM) within rotationally supported galaxies if they trace circular orbits extensively. 
Photometric asymmetries may signify non-circular motions, requiring spectra with dense spatial coverage.
Our integral-field spectroscopy of \targets\ galaxies spanned the mass range of the SAMI Galaxy Survey.
We derived circular speed curves (CSCs) of gas and stars from non-parametric \diskfit fits out to $r\sim$2r$_{\rm e}$. 
For 12/14 with measured H~I profiles, ionized gas and H~I maximum velocities agreed.
We fitted mass-follows-light models to 163 galaxies by approximating the radial starlight profile as nested, very flattened mass homeoids viewed as a \sersic\ form.
Fitting broad-band SEDs to SDSS images gave median stellar mass/light 1.7 assuming a Kroupa IMF vs.\ 2.6 dynamically. 
Two-thirds of the dynamical mass/light measures were consistent with star+remnant IMFs.
One-fifth required upscaled starlight to fit, hence comparable mass of unobserved baryons and/or DM distributed similarly across the SAMI aperture that came to dominate motions as the starlight CSC declined rapidly.
The rest had mass distributed differently from starlight.
Subtracting fits of \sersic\ profiles to 13 VIKING $Z$-band images revealed residual weak bars. 
Near the bar PA, we assessed $m = 2$ streaming velocities, and found deviations usually $<30$ \kms\ from the CSC; three showed no deviation.
Thus, asymmetries rarely influenced our CSCs despite colocated shock-indicating, emission-line flux ratios in more than 2/3.
\end{abstract}


\section{Introduction}
Dark matter (DM) in galaxy discs has been probed both by detailed optical \citep[e.g.][]{Rubin85} and H~I studies \citep[e.g.][]{deBlok08} of dozens of individuals, and in shallower long-slit surveys of a few hundred ranging over environment \citep[e.g.][]{Courteau97}.
These analyses show that possible non-circular motions, uncertain stellar M/L ($\Upsilon_\star$ hereafter), and diverse predictions of DM, make dynamical decomposition indeterminate \citep[e.g.][]{Dutton05}.
In consequence, kinematical models over the optical extent of a galaxy are fitted equally well using priors ranging from almost all DM in diverse radial profiles \citep[e.g.][]{Noordermeer07,deBlok08} to minimal DM to a modified Newtonian dynamics \citep[MOND,][and references therein]{Milgrom83,McGaugh10}.
MOND matches many circular speed curves (CSCs) without DM just by upscaling baryonic (starlight+gas) M/L by separate but constant-with-radius factors for bulge and disc \citep{Sanders96}.
Sometimes ``H~I scaling'' \citep[e.g.][]{Hoekstra01} works: after fitting a ``maximal starlight disc", scaling the observed H~I mass density places enough DM within the extended disc to fit the combined optical+H~I CSC,
and tightens \citep{Pfenniger05} the baryonic Tully-Fisher relation \citep{McGaugh00}.
The required H~I scale factor decreases from $\sim9\times$ in luminous galaxies.
 
Now, multi-headed integral-field feeds to a spectrograph \citep[e.g.][]{Bland11} can map efficiently and uniformly the kinematics of warm plasma and starlight across thousands of galaxies in diverse environments to isolate structural components including bars.
Mass densities mapped by these spectra accelerate well above the MOND critical value $a_o/G$ with $a_o\sim10^{-10}$~m~s$^{-2}$, so Newtonian dynamics suffice.
These uniform datasets allow coordinated estimation of the stellar population/formation history, starlight reddening, and both gas and stellar kinematics.
Perhaps this powerful synergy can better constrain DM possibilities.

Therefore, this paper is a reconnaissance of \targets\ rotationally supported systems from the first quarter of the ongoing SAMI \citep[Sydney-AAO Multi-object integral field spectrograph,][]{Croom12} Galaxy Survey \citep[SGS,][]{Bryant14} of visible-light spectra across the central 15 arcsec diameters of $\sim3400$ galaxies in diverse environments at $0.005\leq z\leq 0.10$.
A goal is to assess if SAMI spatial coverage and sampling can address DM content and bar induced motions, respectively, within $\sim2r_{\rm e}$ of SGS galaxies.
Although the SGS is an optical survey, so relatively dust sensitive, optical stellar populations are less controversial than those in rest-frame NIR surveys \citep[e.g.][and references therein]{Conroy13}.
\citet{Schaefer15} map radial variations of dust attenuation in the SGS; we assumed spatial averages.
Incorporating this refinement will necessitate refitting broad-band photometry, which is beyond our scope.

Section 2 explains how we mapped gas kinematics and derived masses.
In \S3 we report CSCs, and the resulting mass and M/L measured dynamically 
(\MXL hereafter), by imposing consistency with exponentially declining star formation and plausible IMFs ($\Upsilon_\star$ hereafter), and by fitting
stellar population SEDs to $ugriz$-band photometry ($\Upsilon_{\rm P}$ hereafter).
Section 4 discusses our results on disc and bar kinematics, and 
compares these $\Upsilon$s averaged over each galaxy.
The inadequacy of photometric mass to account for the CSC revealed what non-stellar masses must do.
Section 5 concludes.

\begin{figure*}
\centering
\includegraphics[height=0.9\textheight]{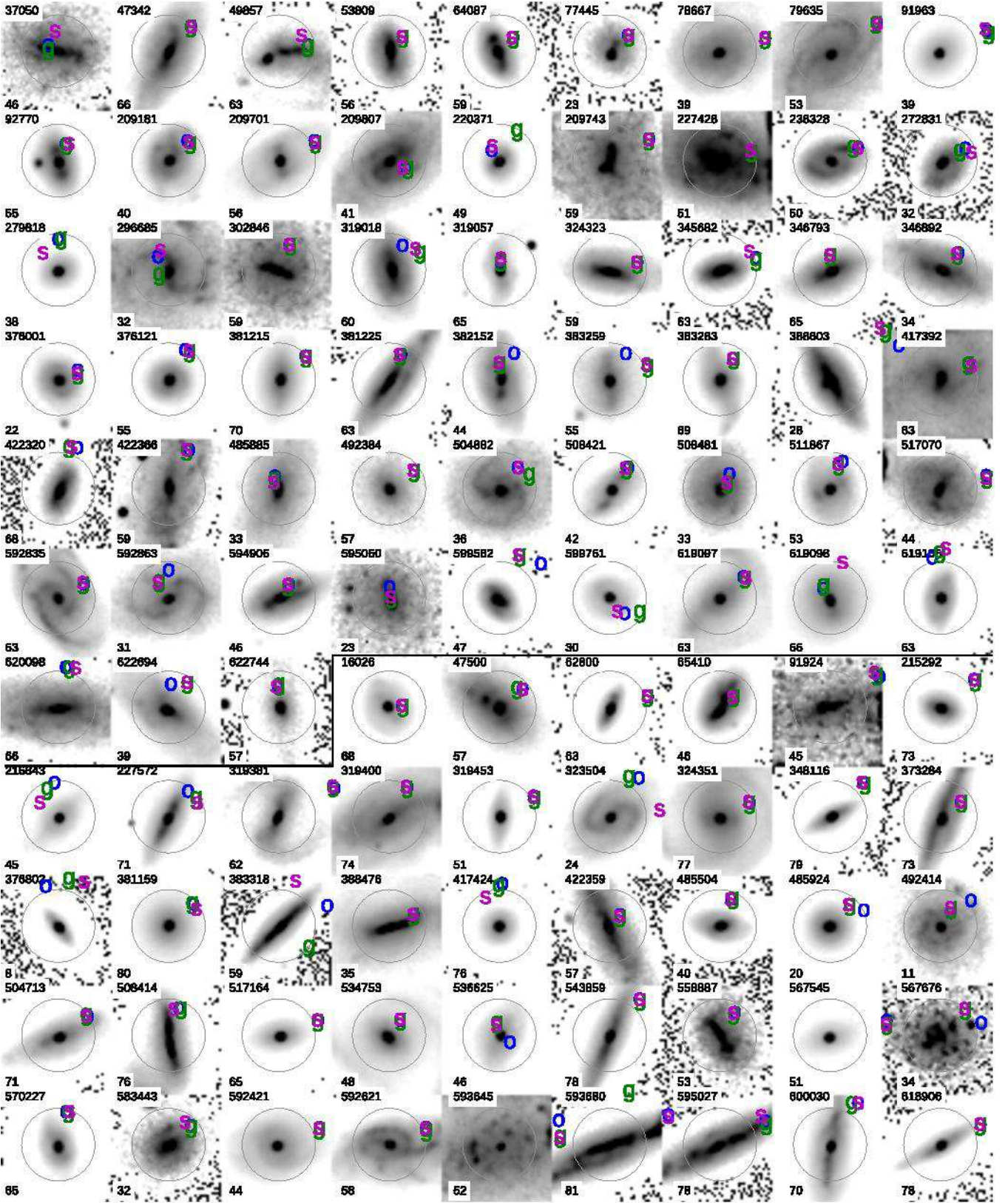}
\protect\caption{\label{fig:ZbandDeconv}VISTA/VIKING survey $Z$-band log-intensity scaled images of some of our GAMA sub-sample, each 22 arcsec on a side with N at top, and having $\sim0.9$ arcsec FWHM after deconvolution using PSF stars with 150 iterations of the Richardson-Lucy algorithm.
SAMI coverage is encircled.
Labeled at \sersic\ $r_{\rm e}$ are major axis PA as blue `o' for photometric, green `g' for gas, and magenta `s' for stars.
The galaxy ID is top left, and its inclination in degrees is bottom left.
Notable are the often very compact bulges. 
Those above the line are in our final sample, those below were rejected for reasons including inclination outside our limits, a dominant bar that confounded a photometric inclination estimate, disorganized motions, or interaction.}
\end{figure*}

\begin{figure*}
\centering
\includegraphics[width=0.69\textwidth]{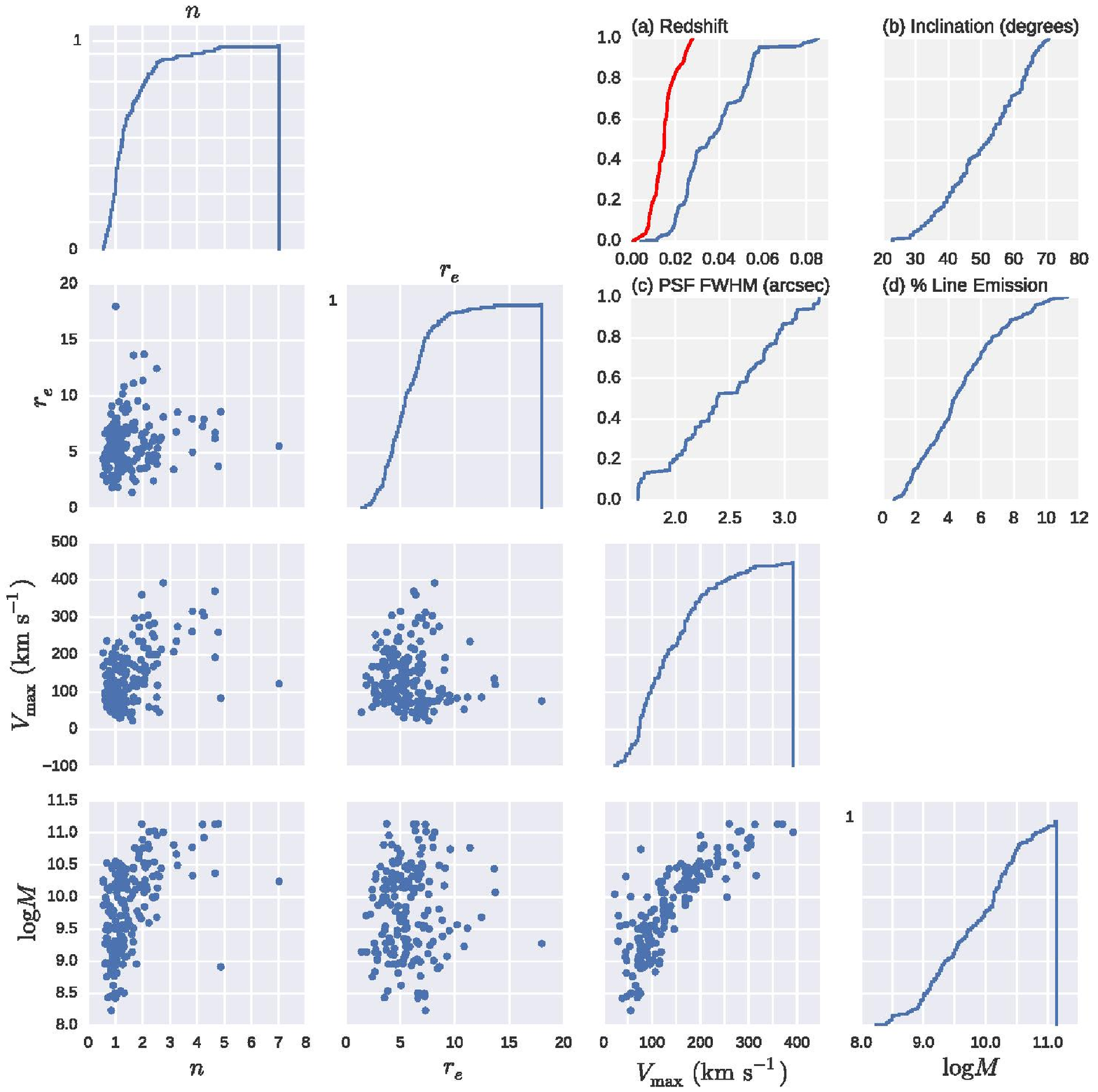}
\protect\caption{\label{fig:sample-properties}Properties of our GAMA sub-sample of 163 galaxies.
Stellar mass estimates come from the GAMA catalogues, and are based on population synthesis fits to optical photometry as described in \citet{Taylor11}.
$r_{\rm e}$ and $n$ come from the single-component \sersic\ fits of \citet{Kelvin13} to SDSS images.
$V_{\rm max}$ comes from our CSC fits.
In red in (a) are the redshifts of the bar sample of \citet{Fred14} from the CALIFA survey \citep{Sanchez12}.}
\end{figure*}

\section{Methods}
From the first $\sim830$ targets observed in the SGS, we selected 344 rotationally supported galaxies having enough gas to map their CSC.
We rejected 8 whose inclination angle to us is too small ($i<20$\arcdeg) to be established reliably by photometry, and those very strongly barred or in obvious interactions.
Finally, we rejected those whose CSC would be smeared excessively by our PSF (\S2.3.1)
because of large inclination ($i>71$ \arcdeg), compact size, or observed in atrocious conditions, leaving 163 SGS GAMA survey sub-sample and 15 ``cluster'' sub-sample galaxies with discs.
Several dozen are in the public SGS Early Data Release \citep{Allen15} and 
span the full mass range of the SGS \citep{Bryant15};
many more cluster discs will be analysed in a paper in preparation.
Fig.\ \ref{fig:ZbandDeconv} shows examples and Fig.\ \ref{fig:sample-properties} summarizes their structural properties.
Notable are compact bulges, evident as median $r$-band \sersic\ shape parameter $n\sim1.4$ (Fig.~\ref{fig:ZbandDeconv}e).

We mapped their CSCs using \diskfit \citep{Sellwood09,Sellwood10}.
For the GAMA survey \citep{Driver04} sub-sample, we had \galfit \citep{Peng10} \sersic\ radial profile fits to SDSS bands \citep{Kelvin12} and VIKING $Z$-band \citep{Lange15} images (e.g.\ Fig.~\ref{fig:ZbandDeconv}).
Subtracting this profile highlighted residual photometric asymmetries to investigate for kinematical disturbances; we estimated gas streaming in 13.
We did not examine environmental influences \citep[e.g.][]{Amran94} because barely 1/3 of the SGS had been observed at the time of writing. 

\subsection{SGS Observations and Data Processing\label{sec:processing}}
SAMI is comprised of 13 integral-field units (IFUs) that are plugged into a custom field plate of 1\arcdeg\ diameter on sky at the corrected prime focus of the 3.9-m aperture Australian Astronomical Telescope (AAT). 
Simultaneously, its 26 single fibres obtain the sky spectrum, while 3 coherent-fibre bundles direct images of field stars to the telescope guider CCD. 
Each IFU has 61 optical fibres each of core diameter 105~\micron\ (1.6 arcsec on sky) and cladded diameter 115~\micron\ \citep{Bryant15}. 
The fibres are arranged in 4 concentric rings that are fused together lightly to obtain 73 percent fill factor over 14.9 arcsec diameter on sky.
The fibres are then separated and routed to the AAOmega bench mounted spectrograph where light is split by a dichroic filter into blue and red beams thence through volume-phase-holographic gratings onto separate E2V CCD $2\times4$ K detectors. 
Blue spectra spanned $\lambda\lambda$370--570 nm (i.e.\ covered the SDSS $g$-band) at resolution $R=\Delta\lambda/\lambda\sim1730$, red spectra spanned $\lambda\lambda$625--735 nm at $R\sim4500$ to approximate SDSS $r$-band for the farther half of our sample.

Please refer to \citet{Bryant15} for SGS target selection and observing procedures, and to \citet{Allen15} and \citet{Sharp2014} for data processing.
Median FWHM of our sample was 2.3 arcsec from Moffat function fits to simultaneously observed stars, providing $\sim40$ independent spatial samples.
Variance and covariance arrays were returned, the latter essential because SAMI spectra are correlated in $0\farcs5\times0\farcs5\times0.1$ nm data cubes.

\subsubsection{Emission-line Processing}
Calibrated cubes were then processed through the \textsc{lzifu} code \citep{Ho14} to isolate and quantify the galaxy emission-line spectrum; we restrict our comments to further custom processing and analysis of its outputs.
We corrected for Galactic attenuation using dust maps from \citet{Schlafly11}, 
then corrected for an attenuating dust screen at the galaxy by assuming that the intrinsic Balmer decrement flux ratio away from an AGN is \ha/\hb=2.86 for case B recombination at $10^4$~K and $n_{\rm e}=100$~cm$^{-3}$. 
Following \citet{Calzetti01}, we obtained luminosity
\begin{equation}
L_{\rm int}(\lambda)=L_{\rm obs}(\lambda)10^{0.4A_\lambda} =
L_{\rm obs}(\lambda)10^{0.4k_\lambda E(B-V)_{\rm stars}}
\end{equation} with the reddening curve band-averaged $k_{\rm r} = 1.16$, $E(B-V)_{\rm stars} = 0.44 E(B-V)_{\rm gas}$, and 
$E(B-V)_{\rm gas}=2.15\log_{10}[$(\ha/\hb)$_{\rm MWcorrected}/2.86]$.
Calzetti shows that this correction recovers the luminosity of star-bursting
systems to a factor of 2 uncertainty, because internal dust is quickly destroyed in the burst to leave an idealized external screen.
We formed the decrement distribution from points with all but the faintest 40 percent of the line-free R\_CONTINUUM \textsc{lzifu} image (approximating $r$-band). 
We corrected for line flux by using the same region; the medians of both of these distributions corrected the entire galaxy.
A future paper will refit \sersic\ profiles to multi-broadband photometry after radial corrections for dust attenuation and line emission.

\subsubsection{Stellar Velocity Processing}
\citet{Fogarty14} detail SGS starlight processing.
We considered only the blue data cube, $\lambda\lambda$370--570 nm at resolution $R=\Delta\lambda/\lambda\sim1730$ to include most important stellar absorbers.
A spectrum of high S/N was made over 2 arcsec diameter centred on the galaxy, to which we fitted 985 MILES \citep{SanchezBlazquez06} stellar spectra using \texttt{ppxf} to adjust the coefficients up to fourth order truncation of the Gauss-Hermite series.
The best composite template was then readjusted with this polynomial to fit the LOSVD of every spaxel in the data cube whose S/N $>5$. 
In the blue data cube this criterion is very close to S/N per \AA.
h$_3$ and h$_4$ often showed just noise for our sample galaxies.

\subsection{Surface Photometry}
We did not separate disc from bulge with different flattening and $r$-band $\Upsilon$, but instead assumed that all starlight of constant $\Upsilon_\star$ is emitted by a flattened homeoid of constant density within each nested shell.
This is an approximation because disc light viewed as a \sersic\ form varies on mass homeoids.
A circular thick disk viewed at inclination $i$ has eccentricity $\epsilon$ related to its intrinsic thickness ratio $q$ by
\begin{equation}
q^{2}=\dfrac{\left(1-\epsilon\right)^{2}-\cos^{2}i}{1-\cos^{2}i}
\end{equation}
We used $q=0.1$ but $q$ from 0.05 to 0.25 produced comparable results, altering the CSC amplitude by $<15$ percent and multiplying the \MXL values in Table 1 column (10) by 0.76 and 1.3, respectively.
$q=0.25$ is appropriate for a dominant ``bulge'', which Fig.~\ref{fig:sample-properties} shows is rare in our sample and confirmed by the few with \sersic\ shape parameter $n>2.5$ in Fig.~\ref{fig:sample-properties}.
The \textsc{sigma} (Structural Investigation of Galaxies via Model Analysis) script of \citet{Kelvin12} calls \textsc{source extractor} \citep{Bertin96}, \textsc{psf extractor} \citep{Bertin13}, and \galfit to fit a radial flux profile with the \sersic\ function of index $n$, shape $b_n = 1.9992n-0.3271$, of surface brightness $\Sigma_{\rm e}$ at effective radius $r_{\rm e}$.
\textsc{sigma} returns the surface brightness in mag arcsec$^{-2}$ $\mu_{\rm e}$ at $r_{\rm e}$, $I_o(n) = e^{b_n} 10^{0.4(21.572+M_{\rm r\odot}-\mu_{\rm e})}$, and the apparent magnitude extrapolated from a few $r_{\rm e}$ out to $10r_{\rm e} \equiv m_{\rm 10re}$ (often $\sim30$ mag arcsec$^{-2}$).
From the last, galaxy absolute magnitude is $M_r = m_{r,10re} - (5\log_{\rm 10}D_{\rm L} + 25+ K_r+A_r)$ for K-correction $K_{\rm r}=1.2z$ \citep{Kelvin13} to $z=0$, dust absorption $A_{\rm r}$, and $D_{\rm L}$ the luminosity distance in Mpc.
Although some of these quantities are covariant, we had only their uncertainties.
We converted to solar luminosities by assuming that solar absolute magnitude in $r$-band $M_{\rm r\odot} = 4.67$.
We subtracted \textsc{lzifu}-derived fluxes of relevant emission-lines for each galaxy from $m_{\rm 10r_e}$.
This line-emission correction and our dust correction $A_r$ 
both used median values over the 40 percent brightest $r$-band continuum.

To compare to dynamical masses, we fitted broad-band SED templates across the $ugriz$-band images from the Sloan Digital Sky survey (SDSS) DR9; please consult \citet{Taylor11} for details on Bayesian procedures and priors also used here.
These fits mapped stellar mass, optical-band starlight weighted ages, and photometric M/L across our sample galaxiess; all these assume solar metallicity.
Comparing radial patterns of dynamical vs.\ stellar mass and age is beyond our scope, so in \S3.4 we just use values averaged over the brightest 40 percent of the $r$-band starlight of each galaxy.
Hereafter, we denote the averaged M/L derived thus as $\Upsilon_{\rm P}$.

As outlined in \S2.5, to explore streaming motions we examined the residual images formed by subtracting the single-\sersic\ fits from SDSS $r$-band and VIKING $Z$-band images.

\subsection{Fitting Disk Velocities}
We now describe how we extracted CSCs by fitting observed gas motions to a disc that approximated the flattened mass homeoids.
Non-circular motions can appear in either a warped disc (an inclination $\phi$-warp or in-plane PA-warp) or flat disc with $m>0$ modal velocity distortions \citep{Staveley90}.
\diskfit handles such motions independently but not in combination.
We first discuss the assumptions of this code, how it accounted for beam smearing, and its use to map rotational and streaming motions within a galaxy.

\subsubsection{Assumptions}
\diskfit fits radial variations of line-of-sight emission-line velocities across a disc all at once with common centre; it does not parametrise velocities over radius nor does it require any asymmetries to be weak.
It is a kinematical -- not hydro-dynamical -- code, so does not address bar or spiral-arm shocks. 
\diskfit bootstraps uncertainties of its model velocities at each spaxel by scrambling fit residuals around each elliptical ring, then adding those to the original data and refitting (see the above references for details); we used 800 resamples at each radius, running these in parallel after updating \diskfit to use modern Fortran coarrays for multi-core efficiency.

Each fitted spaxel $n$ has associated uncertainty $\sigma_{n}$, complicated for SAMI by high spatial covariance with neighbours \citep[][discuss this covariance]{Sharp2014}. 
Covariance is specific to each plug plate of 12 galaxies from seeing variations during the 7 constitutive exposures. 
\textsc{lzifu} reports the variance of its parameter estimates and handles spectral covariance between its multiple velocity components. 
To reduce covariance, we rebinned spectra into $2\times2$ = 1 arcsec$^{2}$ spaxels and used Monte Carlo methods to propagate spatial covariance and variances to a CSC.
To ensure statistical independence in $\sim2$ arcsec FWHM seeing we should have told
\diskfit to ``stride'' across the velocity map at every 2 binned pixels.
However, that would have led to $<5$ independent spaxels from too many elliptical rings, too few for a reliable $\chi^2$ contribution (Eg.\ 4).
We therefore sampled at 1 arcsec.

\begin{figure*}
\centering
\includegraphics[width=0.69\textwidth]{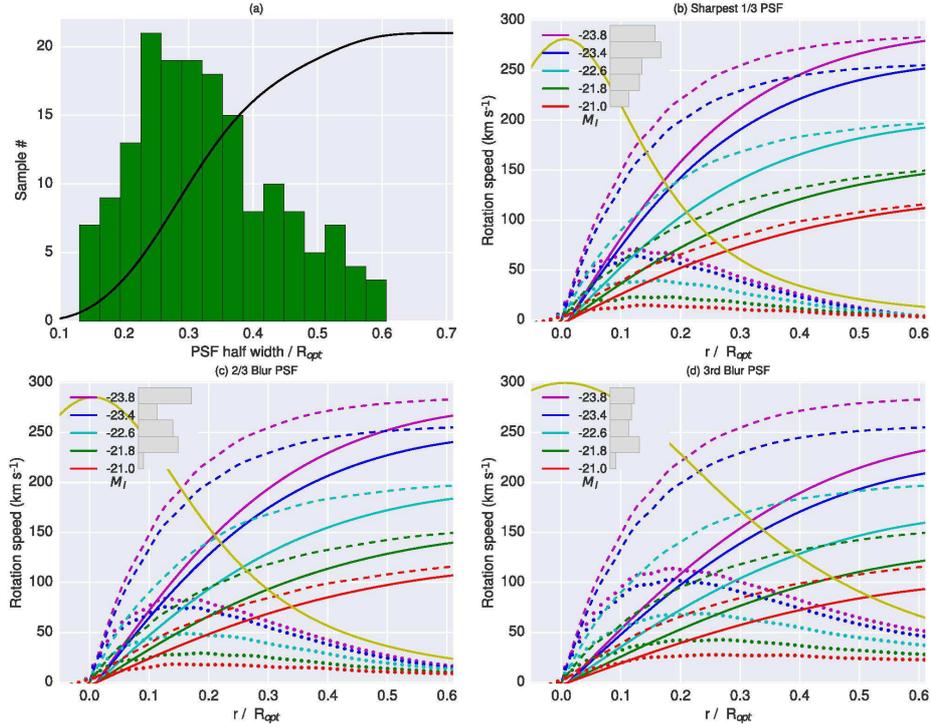}
\protect\caption{\label{fig:beamsmear}How the SAMI PSF smears inner CSCs for our mass range.
(a) Distribution of (Moffat-function PSF Half-Width)/$R_{\rm opt}$. (b)--(d): dashed are the \citet{Catinella06} parametrisation of \textbf{disc only} CSCs for the galaxy absolute magnitude bins M$_I$ shown.
Solid lines: after convolving these with the PSF in yellow at the quartile boundary. 
Dotted: differences.
The grey histograms show the relative number of sample galaxies in each M$_{\rm r}$ bin, approximating M$_{\rm I}$ as described in the text.
}
\end{figure*}

``Beam smearing'' can blur velocity gradients to distort the inner rise of the CSC \citep[e.g.][]{Sofue99} where stellar bars may influence velocities and secular evolution \citep{Kormendy04}.
It manifests as a broad uncertainty on each CSC point from light smeared into the fitting ellipse; the defect increases with galaxy inclination.

To assess its impact, we used the CSC parametrisation of \citet{Catinella06} in terms of galaxy $M_{\rm I}$ and
$R_{\rm opt}$ (83 percent of the $R$-band light is encircled on sky). 
We approximated the former from the \galfit of \citet{Kelvin12} to the SDSS $r$-band image (FWHM $\la1.5$ arcsec)s, adjusted by $(r-i)=0.40\pm0.05$ dex (originally from \citet{Blanton03}, revised by \citet{Kelvin12}) appropriate for blue-sequence galaxies.
Replacing the latter, we formed a more sensitive metric of smear by first taking
the median of the $r$-band \sersic\ $r_{\rm e}$ as projected around the ellipse.
We then divided that into the half-width at half-maximum of the Moffat-function fit to the 
contemporaneous SAMI PSF to rank the sample from least to most blurred, see column (8) of Table 1.
Panels (b)--(d) of Fig.~\ref{fig:beamsmear} show, for the $M_{\rm I}$ from the \galfit to each galaxy, the \citet{Catinella06} parametrisation smeared with various quartile seeing.
The linear rise of CSC of more massive galaxies ($v_{\rm max}>200$ \kms) was undistorted only in the sharpest quartile; for less massive ones, up to median blur was acceptable.

To de-smear, \diskfit smoothes the fit model by adjusting the fractional
contribution of the nearest 11 spaxels to a spaxel in the nominal fitting ellipse.
This tactic succeeds only when the CSC is sampled less often than twice the seeing FWHM. 
For most of our datacubes this would require $>4$ arcsec sampling, hence leaving only $1-2$ points plus $v_{\rm sys}$ to trace the CSC.
Instead, we explored the effectiveness of partial correction by tracking how CSC points and their uncertainties adjusted as
smear correction progressed -- and perhaps converged -- from 0 to 0.75 to 1 arcsec FWHM.
We found that most bootstrap uncertainties shrank considerably with only the 0\farcs75 FWHM corrective redistribution, to deliver an informative CSC at 1.5 arcsec increments.
The corrected distribution at each radius was centred in the upper half of the original distribution, as expected.

The \textsc{lzifu} multi-Gaussian parametrisation of each spectrum accounts for stellar absorption that might shift the velocity centroid of line emission.
We fitted to gas velocities only within an ellipse centred on the nucleus whose major axis extent and PA was set by the \ha\ map, and whose minor axis extent was just the projected major axis extent. 
We omitted nearly face-on or highly inclined galaxies, considering only 20\arcdeg $< i <$ 71\arcdeg.
We generally fixed inclination to that derived by \textsc{source extractor} from the SDSS $r$-band image (see \S2.2); the \textsc{sigma} code did not output \textsc{source extractor} inclination uncertainties, so we assumed 5 percent of the mean for its $\sigma$.
A warped or non-circular disc will give a bad prior inclination, most strongly influencing the CSC of near-face-on modeled plane discs. 
Inclination warps seem to be rare for optical discs, but often start beyond in H~I \citep{Sancisi76}). 
PA-twists are less well studied but more common due to bar/oval streaming.
We found that \diskfit was unreliable on PA-twisted SGS discs, so those were omitted
when we defined our sample.

We fitted two sets of velocities: 1) just the one-Gaussian \textsc{lzifu} model, which is robust but can miss flux in the wings of emission-line profiles; 2) full non-Gaussian profiles obtained by synthesizing the \ha\ emission-line profile from all \textsc{lzifu} components. 
We mapped uncertainties for 2) by random sampling the \textsc{lzifu} Gaussian uncertainty distributions of the fitted multi-component velocities and their dispersions, \ha\ fluxes, and various estimates of the Balmer $\alpha$ absorption correction provided by different spectral templates, each time recomputing the flux weighted median velocity. 
The median of the resulting distribution of medians was the adopted velocity at that spaxel, and its $\pm34$ percent spread averaged around median was our error weight for the CSC fit.
Our fits ignored outliers in the deprojected ellipse that deviated by $>50$ \kms from neighbours.

We fitted two models with smear correction 0.75 arcsec FWHM: 1) for the GAMA sub-sample with its photometric Gaussian prior on inclination, allow the PA to vary; 2) also vary inclination from the \galfit model estimate \citep{Kelvin12} of the \sersic\ profile; the centre was always allowed to vary by $\pm0.5$ arcsec from its photometric prior value.
We found that tactic 2) never improved the fit.
Galaxies in our SGS ``cluster'' sub-sample lacked GAMA priors, so fits always varied disc inclination and PA.
These extra freedoms increased variance of the CSC fit, so we imposed coincident kinematic and photometric centres to stabilise those fits. 
\textsc{sigma} fits of the full SGS ``cluster" sub-sample photometry are underway; a paper in preparation will present their CSCs and mass models.

\citet{Ho14} demonstrate that 
\textsc{lzifu} maps emission-lines whose flux ratios diagnose shock velocities and the shock fraction of the total shock+photoionised emission. 
Shock models are parametrised by the ionisation of pre-shocked gas, shock velocity, and magnetic pressure.
We used the \textsc{mappings3} code \citep{Dopita03} inputs of \citet{Ho14} that were bounded for the SGS by combining [\ion{O}{iii}]/\hb\ with [\ion{S}{ii}]/\ha\ or [\ion{N}{ii}]/\ha\ shock diagnostics; these choices ranged over the ratios evident in our sample.
We considered spaxels where shocks contributed at least 40 percent of the total emission.
By correlating the shock maps with the map of $m=2$ residuals, we assessed if shock velocities were consistent with kinematical deviations. 
We did not tune model pre-shock ionisation or magnetic pressure.

\begin{figure}
\centering
\includegraphics[scale=0.37]{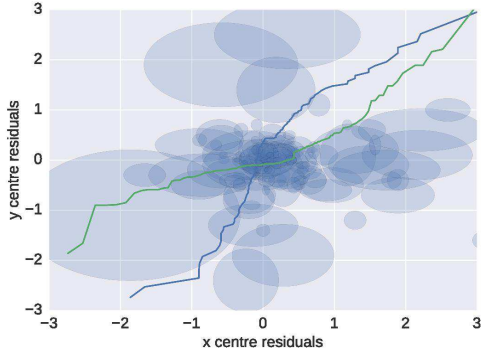}
\protect\caption{\label{fig:centers}Difference between some photometric nominal and fitted kinematical disc centres; unit is 0.5 arcsec pixel and uncertainties are $\pm1\sigma$.
The cumulative histograms of the differences are also shown, starting from the lower left corner.}
\end{figure}

\subsubsection{Axisymmetric Fits}
For these, $V_{\rm 2,t}(r)=V_{\rm 2,r}(r)=0~~\forall~~r$. 
Fig.~\ref{fig:centers} shows the displacements between rotational and WCS centres.
As mentioned, for the GAMA sub-sample we fixed $i$ to the value derived from the single-\sersic\ \galfit to SDSS $r$-band photometry \citep{Kelvin12}.

\subsubsection{Fitting Procedure}
To derive $V_{\rm t}(r),$ $V_{\rm 2,t}(r),$ $V_{\rm 2,r}(r),$ $\phi_{\rm b}$, $V_{\rm sys}$, $i,$ and PA $\theta$ from the major axis, \diskfit fits observed gas velocities to
\begin{equation}
\begin{split}
V_{\rm model} & =V_{\rm sys}+[V_{\rm t}(r)\cos\theta-V_{\rm 2,t}(r)\cos m(\theta- 
\phi_{\rm b})\cos\theta \\
& -V_{\rm 2,r}(r)\sin m(\theta-\phi_{\rm b})\sin\theta]\sin i
\end{split}
\end{equation}
At $N$ sky positions $(x,y)$ it minimizes
\begin{equation}
\chi^{2}=\sum_{n=1}^{N}(\frac{V_{\rm obs}(x,y)-\sum_{k=1}^{N}w_{k,n}V_{k}}{\sigma_{n}})
\end{equation}
\noindent Velocity uncertainties are taken uncorrelated of form $\sigma_{n}(x,y)=\sqrt{\Delta_{\rm D}(x,y)^{2}+\Delta_{\rm ISM}^{2}}$
with $\Delta_{\rm D}$ our estimated velocity uncertainty described above.
$\Delta_{\rm ISM}$ is ``ISM turbulence'', increased \textit{ad hoc} from 0 until the reduced $\chi^{2}$ was approximately normalized; see Tables 1 and 2.
\citet{Sellwood09}, \citet{Sellwood10}, and the code documentation provide details.
One anticipates more dispersion near strong bars, e.g.\ $\sigma_{\rm z,bar}\sim30$ \kms\ for young bars and $\sim100$ \kms\ for evolved ones \citep{Gadotti05}, but we had excluded atypical, strong photometric bars \textit{a priori}.

\subsubsection{Non-axisymmetric Fits to Streaming Motions}
After beam correction, inner points whose velocities deviated from the photometrically motivated CSC suggested the need to evaluate non-circular/decentred motions.
Such motions affect CSC values and uncertainties even beyond the distortion because the galaxy centre is adjusted to minimize fit residuals across the disc.
Consistent deviation across several rings is needed for reliability, hindering application of \diskfit to SAMI data cubes with their at most 50 independent spatial points.

For 13 galaxies, velocity residuals never shrank after adding to the model an $m=2$ asymmetry whose fixed major axis PA extended slightly beyond the photometric single-\sersic-fit residual.
To these we first fitted with a variable bar PA, then explored how $\chi^2$ varied adjacent to the optimal value.
If $\chi^2$ changed negligibly, we simply aligned kinematical and photometric bars.
$m=0$ and $m=2$ components are degenerate, hence fits are unreliable, when the two PAs differ by $\sim0^\circ$ or $\sim90^\circ$.
Using an $m=1$ mode likewise did not shrink fit residuals.
The Appendix discusses the non-axisymmetric systems.

\subsection{Estimating Dynamical M/L and Mass}
CSCs across the optical extent of some galaxies can be fit well with variable $\Upsilon(r)$ \citep[VML, e.g.][]{Takamiya00}, reaching $\la10\Upsilon_\star$.
To explore the mass distribution that drives the CSC, we used SDSS DR9 $r$-band images
because of reduced S/N and uncertain $\Upsilon_\star$ at longer wavelengths.
We sought the fraction of a CSC generated by stars to assess as yet unobserved baryons and, ultimately, DM by any deficit between model and data.
We denote by \MXL the maximum fitted value unconstrained by IMF considerations.
It merely scaled starlight to CSC by asserting constant $\Upsilon_r$ (CML hereafter) to see how much DM might be distributed like starlight.
The few fits that still failed must have VML if their values were $\sim\Upsilon_\star$,
or mass that does not follow light.

We integrated numerically the mass density contributed at each radius from the fit to the \sersic\ profile
\begin{equation}
I_S(x) = I_o(n)\exp{[-(\dfrac{x}{r_o})^{1/n}]}
\end{equation}
to obtain the CSC$(r)^2$ 
\begin{equation}
V^2(r) = C\int\limits_{0}^r \dfrac{m^2dm}{\sqrt {{r^2}-{m^2}{\varepsilon ^2}}}{\int\limits_{m}^\infty \Upsilon_{\rm r}(x)(\dfrac{x}{r_o})^{1/n-1} \dfrac{{e^{-(x/{r_o})^{1/n}}dx}}{\sqrt {x^2-m^2} }}
\end{equation}
with $C =\dfrac{4Gq I_o}{r_o n}\sqrt {{{\sin }^2}i + {{\cos }^2}i/q^2}$.
We used either CML $\Upsilon(r)=1$ or VML 
\begin{equation}
\Upsilon(r) = \Upsilon_{\rm r,1} \exp[-\alpha(s)\{(\dfrac{r}{r_{\rm e}})^s-1\}].
\end{equation}

Monte Carlo methods propagated uncertainty distributions of assumed $A_r$, and the \textsc{lzifu} velocity and line-emission luminosity corrections, to the CSC.
These uncorrelated errors weighted the linear least-squares fit (\texttt{Mathematica v10 LinearModelFit}) to bound $\Upsilon$ uncertainties.

\onecolumn
\footnotesize
\begin{center}
\tablefirsthead{
\multicolumn{11}{c}{\textbf{Table 1}} \\
\multicolumn{11}{c}{\diskfit Results for Some of the SGS GAMA-Survey Sub-sample\label{tab:fitresults}} \\
\toprule
GAMA & \multicolumn{1}{c}{$z$ Fitted} & Line & \multicolumn{1}{c}{$\Delta$ISM} & \multicolumn{1}{c}{Inclination} & \multicolumn{1}{c}{PA} & \multicolumn{1}{c}{$\chi^2$ (DOF)} & V$_{\rm max}$ & \multicolumn{1}{c}{r$_{\rm max}$} & \multicolumn{1}{l}{} & \multicolumn{1}{c}{$\Upsilon_\times \times L_{\rm r}$}\\
ID & & (\%) & \multicolumn{1}{c}{(km~s$^{-1}$)} &   \multicolumn{1}{c}{(deg.)} & \multicolumn{1}{c}{(deg.)} & \multicolumn{1}{c}{Fit} & (km~s$^{-1}$) & \multicolumn{1}{c}{($r_e$)} & \multicolumn{1}{r}{$\Upsilon_\times,~\Upsilon_{\rm P}$} &   \multicolumn{1}{c}{($10^9$ M$_\odot$)} \\
(1) &   \multicolumn{1}{c}{(2)} & (3) & \multicolumn{1}{c}{(4)} &   \multicolumn{1}{c}{(5)} & \multicolumn{1}{c}{(6)} & \multicolumn{1}{c}{(7)} & \multicolumn{1}{c}{(8)} & \multicolumn{1}{c}{(9)} & (10) &   \multicolumn{1}{c}{(11)} \\
\toprule
}
\tablehead{
\multicolumn{11}{l}{continued ...} \\
\toprule
GAMA & \multicolumn{1}{c}{$z$ Fitted} & Line & \multicolumn{1}{c}{$\Delta$ISM} & \multicolumn{1}{c}{Inclination} & \multicolumn{1}{c}{PA} & \multicolumn{1}{c}{$\chi^2$ (DOF)} & V$_{\rm max}$ & \multicolumn{1}{c}{r$_{\rm max}$} & \multicolumn{1}{l}{} & \multicolumn{1}{c}{$\Upsilon_\times \times L_{\rm r}$}\\
ID & & (\%) & \multicolumn{1}{c}{(km~s$^{-1}$)} &   \multicolumn{1}{c}{(deg.)} & \multicolumn{1}{c}{(deg.)} & \multicolumn{1}{c}{Fit} & (km~s$^{-1}$) & \multicolumn{1}{c}{($r_e$)} & \multicolumn{1}{r}{$\Upsilon_\times,~\Upsilon_{\rm P}$} &   \multicolumn{1}{c}{($10^9$ M$_\odot$)} \\
(1) &   \multicolumn{1}{c}{(2)} & (3) & \multicolumn{1}{c}{(4)} &   \multicolumn{1}{c}{(5)} & \multicolumn{1}{c}{(6)} & \multicolumn{1}{c}{(7)} & \multicolumn{1}{c}{(8)} & \multicolumn{1}{c}{(9)} & (10) &   \multicolumn{1}{c}{(11)} \\
\toprule
}
\tablelasttail{
\toprule
\multicolumn{11}{l}{(3) \% of $r$-band flux from line emission. Codes: \dag\ has H I (Fig.\ 8); $\star$ = extended shock w/ asymmetric starlight; $\times$ = extended shock,} \\
\multicolumn{11}{l}{
symmetric starlight; $\circ$ = compact shock at centre; $\textit{a}$ = no shock but asymmetric starlight; otherwise no shock, symmetric starlight.} \\
\multicolumn{11}{l}{(4) ISM dispersion to normalize \textsc{diskfit} reduced $\chi^2$.} \\
\multicolumn{11}{l}{(5) Inclination is fixed at GAMA Survey prior.} \\
\multicolumn{11}{l}{(6) \textsc{diskfit} evaluated PA is east from N, bootstrapped error.} \\
\multicolumn{11}{l}{(7) Reduced $\chi^2$ of fit (degrees of freedom).} \\
\multicolumn{11}{l}{(8) 1 = sharpest quartile, 2 = to median, etc.} \\
\multicolumn{11}{l}{(9) Radius of maximum velocity within the SAMI aperture.} \\
\multicolumn{11}{l}{(10) Dynamical $\Upsilon_\times$, broad-band colour fitted $\Upsilon_{\rm P}$ corrected for line emission and dust. $\ddagger$ \MXL fit failed because mass does not follow} \\
\multicolumn{11}{l}{light.
Changing the intrinsic flattening from $q=0.1$ to $q=0.05$ or 0.25 would multiply the listed $\Upsilon_\times$ by 0.76 or 1.3, respectively.} \\
\multicolumn{11}{l}{(11) Dynamical mass from (10).} \\
}
\begin{supertabular}{rPcPcpllPcrP}
8353 & 0.019623 & 9$\star$ & 1. & 50.9 & 161.8!0.3 & 1.3 (161)& 70 (3)& 1.7 & 1.1, 1.5& 2.9 \\
15165 & 0.077651 & 1$\times$ & 0.5 & 28.4 & 332.7!0.1 & 0.62 (160)& 392 (2)& 1.2 & 4., 2.& 115.6 \\
15218 & 0.025604 & 5$\star$ & 0.5 & 56.6 & 324.6!0.1 & 1.23 (149)& 82 (1)& 1. & 2.9, 1.5& 2.9 \\
23117 & 0.081989 & 2$\star$ & 1. & 43.9 & 37.1!0.3 & 2.43 (148)& 315 (1)& 1. & 1.5$\ddagger$, 1.9& 118.4 \\
23565 & 0.037233 & 2$\star$ & 2. & 56.6 & 29.8!0.1 & 1.17 (153)& 193 (3)& 1.9 & 1.4, 1.9& 23.9 \\
30890 & 0.019945 & 5$\star$ & 2. & 55.9 & 289.3!0.1 & 1.49 (177)& 129 (1)& 1.4 & 2.1, 1.7& 8.2 \\
32362 & 0.019052 & 2$a\dagger$ & 2. & 55.9 & 316.4!0.1 & 1.29 (186)& 156 (1)& 0.8 & 1.5$\ddagger$, 1.9& 27.7 \\
37064 & 0.055354 & 2$\circ$ & 1. & 42.3 & 126.7!0.1 & 1.78 (187)& 270 (2)& 1.8 & 3., 1.8& 47.4 \\
41144 & 0.029716 & 5 & 2. & 53.8 & 149.7!0.1 & 1.08 (166)& 190 (2)& 1.3 & 1.6, 1.8& 29.5 \\
47342 & 0.024147 & 4 & 2. & 66.4 & 334.2!0.1 & 1.1 (138)& 148 (1)& 1.7 & 1.9, 1.9& 10.6 \\
49857 & 0.043516 & 10 & 2. & 63.3 & 93.3!0.1 & 1.17 (124)& 75 (3)& 2.3 & 2.1, 1.4& 3.0 \\
53977 & 0.048448 & 9$\star$ & 2. & 36.9 & 182.!0.1 & 1.09 (144)& 124 (2)& 2. & 1.5, 1.5& 15.7 \\
55537 & 0.078764 & 2$\star$ & 4. & 63.9 & 0.8!0.1 & 1.48 (93)& 240 (3)& 1.6 & 4.9, 1.8& 31.4 \\
55905 & 0.020971 & 6$\times$ & 0.5 & 62. & 192.!0.1 & 0.87 (116)& 77 (1)& 1.1 & 5.5, 1.5& 1.5 \\
56064 & 0.04038 & 2 & 1. & 46.4 & 333.7!0.2 & 1.88 (179)& 166 (2)& 1.3 & 1.3, 1.9& 34.4 \\
56183 & 0.039097 & 8$\star$ & 2. & 39.6 & 167.5!0.3 & 1.14 (94)& 78 (3)& 1.7 & 1., 1.5& 6.5 \\
64087 & 0.055367 & 6$\star$ & 7. & 59.3 & 200.6!0.1 & 0.86 (66)& 229 (3)& 2.2 & 1.9, 1.9& 24.4 \\
65278 & 0.042762 & 4$\times$ & 0. & 69.5 & 44.9!0.1 & 0.28 (66)& 84 (2)& 1.7 & 3.8, 1.6& 2.0 \\
65406 & 0.04307 & 2$\star$ & 4. & 31.8 & 96.6!0.1 .& 3.2 (172)& 394 (1)& 0.9 & 5.4, 2. &57.4 \\
69653 & 0.018423 & 7$\times$ & 2. & 53.1 & 138.9!0.6 & 1.26 (81)& 58 (1)& 0.9 & 3.5, 1.3& 1.1 \\
71382 & 0.021357 & 6$\circ$ & 0.5 & 35.9 & 12.2!0.1 & 0.72 (106)& 94 (2)& 1.5 & 3.6, 1.6& 1.4 \\
77754 & 0.053292 & 5$\star$ & 3. & 55.9 & 338.6!0.1 & 0.89 (174)& 211 (2)& 1.5 & 2.$\ddagger$, 1.7& 44.6 \\
78667 & 0.055065 & 4$\dagger\circ$ & 0.5 & 38.7 & 111.8!0.1 & 1.4 (184)& 167 (1)& 1.5 & 3., 1.6& 23.9 \\
78921 & 0.02992 & 4$\circ$ & 1. & 56.6 & 69.3!0.1 & 0.9 (74)& 95 (1)& 0.7 & 6.3, 1.6& 3.4 \\
79635 & 0.040279 & 2$\dagger\star$ & 1. & 53.1 & 318.2!0.1 & 2.89 (186)& 203 (1)& 1.2 & 3.8, 1.8& 35.5 \\
79771 & 0.042408 & 5$\times$ & 0.5 & 45.6 & 345.8!0.2 & 0.73 (122)& 96 (3)& 1.6 & 6.3$\ddagger$, 1.4& 1.7 \\
84107 & 0.028832 & 10$\star$ & 2. & 39.6 & 186.6!0.1 & 1.72 (183)& 130 (1)& 1.8 & 2., 1.4& 9.5 \\
84677 & 0.047406 & 7$\star$ & 7. & 62.6 & 334.2!0.1 & 0.86 (127)& 225 (2)& 1.5 & 2.7, 1.8& 41.2 \\
85481 & 0.020062 & 3$\times$ & 0.5 & 53.8 & 221.8!0.2 & 0.52 (141)& 96 (3)& 2.1 & 2.3$\ddagger$, 1.6& 1.3 \\
91963 & 0.050017 & 1$\circ$ & 1. & 38.7 & 310.5!0.1 & 0.96 (90)& 293 (1)& 0.8 & 3.8, 2.& 66.8 \\
99349 & 0.019791 & 4$\star$ & 6. & 67.7 & 52.3!0.2 & 1.83 (112)& 172 (1)& 1. & 2., 1.9& 18.2 \\
99393 & 0.020235 & 6$\star$ & 2. & 51.7 & 334.!0.1 & 1.58 (158)& 120 (1)& 1.8 & 1.2$\ddagger$, 1.8& 4.9 \\
100162 & 0.025667 & 3$\star$ & 2. & 60. & 232.5!0.2 & 1.14 (64)& 79 (3)& 2.2 & 1.2, 1.6& 2.5 \\
100192 & 0.023995 & 8$\star$ & 2. & 23.1 & 45.3!0.1 & 0.85 (109)& 102 (1)& 1.1 & 3.2, 1.6& 3.7 \\
105962 & 0.025505 & 6$\star$ & 0.5 & 50.2 & 219.3!0.1 & 0.52 (157)& 77 (2)& 1.5 & 2.3, 1.5& 1.8 \\
106042 & 0.026053 & 9 & 2. & 36.9 & 97.2!0.2 & 1.29 (105)& 120 (1)& 0.4 & 1.2, 1.5& 30.8 \\
106331 & 0.035944 & 7$\star$ & 2. & 55.9 & 157.1!0.3 & 1.26 (152)& 96 (1)& 1.3 & 2.8, 1.5& 7.0 \\
106717 & 0.025704 & 10$a$ & 3. & 45.6 & 109.4!0.1 & 1.17 (139)& 184 (1)& 1.4 & 1.7, 1.6& 20.6 \\
137789 & 0.019107 & 4$\times$ & 0.5 & 45.6 & 128.7!0.6 & 0.34 (121)& 124 (2)& 2.5 & 5.3, 1.6& 0.4 \\
138066 & 0.035207 & 4$\times$ & 0.5 & 53.8 & 270.2!6.3 & 0.55 (163)& 121 (3)& 1.9 & 4., 1.8& 3.2 \\
143814 & 0.019863 & 12$\star$ & 1. & 67. & 107.3!0.1 & 1.8 (131)& 73 (2)& 1.6 & 1., 1.4 &3.9 \\
144402 & 0.035557 & 8$\star$ & 10. & 49.5 & 45.1!0.2 & 1.8 (137)& 201 (2)& 1.8 & 1.6, 1.7& 24.5 \\
178481 & 0.025598 & 4$\times$ & 0.5 & 44.8 & 43.5!0.1 & 0.19 (141)& 49 (1)& 0.9 & 1.5, 1.5& 2.3 \\
178578 & 0.019921 & 4 & 0.5 & 62.6 & 79.4!0.2 & 0.34 (90)& 41 (2)& 1.1 & 2.3, 1.5& 0.5 \\
184281 & 0.019903 & 6$a$ & 0.5 & 64.5 & 178.2!0.6 & 0.45 (102)& 75 (2)& 1.3 & 7.4, 1.5& 0.7 \\
184415 & 0.028315 & 6$a$ & 1. & 40.5 & 88.!0.1 & 1.73 (139)& 110 (3)& 2.1 & 1.4, 1.6& 5.6 \\
185252 & 0.021583 & 9$\times$ & 0.5 & 52.4 & 328.9!0.8 & 0.44 (117)& 58 (1)& 1. & 5.7$\ddagger$, 1.5& 0.7 \\
185291 & 0.021652 & 7$\times$ & 0.5 & 53.8 & 144.2!1.6 & 0.75 (77)& 50 (1)& 1.2 & 1.7, 1.7& 0.9 \\
185532 & 0.020068 & 3$\circ$ & 0.5 & 28.4 & 76.1!0.1 & 1.33 (157)& 96 (1)& 1. & 6.1, 1.6& 1.9 \\
198591 & 0.0117 & 9 & 2. & 48.7 & 241.7!0.1 & 1.35 (167)& 94 (1)& 0.9 & 1.8, 1.7& 3.8 \\
198771 & 0.011684 & 4$\star$ & 2. & 70.7 & 98.1!0.1 & 1.31 (98)& 86 (1)& 0.5 & 4., 1.4 &4.8 \\
204983 & 0.054962 & 1$\star$ & 0.5 & 54.5 & 146.1!0.1 & 0.79 (152)& 255 (3)& 1.9 & 2.9, 1.9& 39.4 \\
205097 & 0.055574 & 2$\star$ & 0.5 & 57.3 & 275.1!0.1 & 0.46 (135)& 195 (3)& 1.7 & 2.9, 1.9& 21.7 \\
209181 & 0.058247 & 8$\star$ & 3. & 39.6 & 324.4!0.1 & 1.42 (197)& 186 (2)& 2.1 & 1.5, 1.5& 33.8 \\
209414 & 0.025703 & 5$\times$ & 1. & 56.6 & 6.1!0.3 & 1.36 (135)& 86 (1)& 1.3 & 5.1, 1.6& 1.6 \\
209698 & 0.028588 & 5$a$ & 1. & 59.3 & 312.4!0.8 & 1.76 (55)& 49 (2)& 1. & 0.1, 1.7& 25.3 \\
209701 & 0.053303 & 1$\star$ & 0.5 & 55.9 & 300.7!0.1 & 2.84 (146)& 297 (2)& 1.3 & 5.2, 1.9& 44.0 \\
209743 & 0.040693 & 4$\dagger$$a$ & 2. & 58.7 & 130.6!0.1 & 1.31 (113)& 173 (2)& 1.1 & 3.2, 1.7& 21.8 \\
209807 & 0.053893 & 4$\star$ & 5. & 41.4 & 165.9!0.1 & 1.3 (151)& 209 (3)& 2.2 & 0.8$\ddagger$, 1.9& 62.6 \\
210660 & 0.016896 & 3$\star$ & 8. & 57.3 & 143.8!0.2 & 0.67 (113)& 128 (2)& 1.4 & 1.1$\ddagger$, 1.8& 8.1 \\
210781 & 0.055356 & 2$a$ & 1. & 40.5 & 89.9!23. & 1.01 (148)& 170 (2)& 1.7 & 2.3, 1.8& 21.3 \\
216843 & 0.023893 & 6$\dagger\star$ & 2. & 44.8 & 224.2!0.1 & 1.74 (93)& 72 (1)& 0.8 & 4.1, 1.6& 2.5 \\
220332 & 0.020281 & 2$\star$ & 0.5 & 68.9 & 95.9!0.1 & 0.47 (73)& 89 (2)& 1. & 1.9, 1.9 & 3.6 \\
220371 & 0.020275 & 4$a$ & 1. & 49.5 & 244.6!0.1 & 1.27 (177)& 91 (1)& 1.5 & 2.3, 1.8& 3.5 \\
220439 & 0.019453 & 9$a$ & 2. & 34.9 & 20.1!0.1 & 0.84 (187)& 129 (2)& 1.6 & 2.5, 1.6& 4.7 \\
220750 & 0.020998 & 5$a$ & 0.5 & 45.6 & 310.2!0.8 & 0.63 (87)& 69 (2)& 1.2 & 4.2, 1.5& 0.8 \\
221375 & 0.027765 & 6$\star$ & 2. & 62. & 39.7!0.1 & 1.46 (129)& 113 (1)& 1.1 & 1.8, 1.6& 11.1 \\
227371 & 0.024828 & 6$a$ & 0.5 & 65.8 & 255.6!0.4 & 0.22 (139)& 78 (2)& 1.5 & 7.6, 1.5& 0.7 \\
227673 & 0.025844 & 7 & 2. & 29.5 & 140.6!0.2 & 1.12 (138)& 102 (2)& 1.6 & 2.6, 1.5& 2.8 \\
227970 & 0.054054 & 6$\star$ & 3. & 40.5 & 204.3!0.1 & 0.84 (156)& 185 (2)& 1.7 & 2.6, 1.6& 24.7 \\
238358 & 0.05433 & 4 & 10. & 41.4 & 293.2!0.1 & 1.03 (96)& 322 (1)& 0.8 & 2.9, 1.8& 94.7 \\
238395 & 0.024997 & 9$\dagger\star$ & 4. & 34.9 & 66.4!0.1 & 1.43 (162)& 118 (3)& 1.8 & 1.$\ddagger$, 1.6& 10.0 \\
239109 & 0.085402 & 1$\circ$ & 2. & 57.3 & 247.4!0.1 & 1.53 (115)& 361 (1)& 1.2 & 4.2, 1.9& 134.0 \\
250277 & 0.058051 & 6$\star$ & 0.5 & 44.8 & 321.3!2.2 & 0.85 (69)& 56 (2)& 1.3 & 0.4, 1.6& 16.3 \\
272667 & 0.022519 & 3$\star$ & 1. & 43.1 & 268.6!0.1 & 1.18 (183)& 83 (1)& 1.2 & 1.9, 1.7& 3.9 \\
273092 & 0.037521 & 4$a$ & 0.5 & 41.4 & 64.!0.4 & 0.35 (145)& 25 (1)& 1. & 0.1, 1.6& 14.2 \\
273296 & 0.020885 & 6$\star$ & 2. & 57.3 & 348.9!0.2 & 1.53 (141)& 91 (1)& 0.8 & 2.2, 1.5& 6.8 \\
273952 & 0.026862 & 6$\times$ & 0.5 & 48.7 & 341.!0.4 & 1.31 (127)& 43 (1)& 1.1 & 0.9, 1.6& 2.7 \\
278548 & 0.043219 & 4$a$ & 5. & 62.6 & 45.1!0.5 & 1.16 (97)& 113 (1)& 1. & 0.9, 1.7& 22.6 \\
278760 & 0.041235 & 9$\star$ & 2. & 60.7 & 279.!0.1 & 1.12 (100)& 137 (1)& 1.6 & 2.6, 1.6& 9.5 \\
278812 & 0.041702 & 8$\star$ & 3. & 65.8 & 223.6!0.1 & 0.81 (87)& 112 (1)& 1.6 & 2.5, 1.5& 6.4 \\
279818 & 0.027289 & 7$\dagger\star$ & 2. & 37.8 & 194.!0.3 & 1.44 (153)& 44 (1)& 1. & 0.5, 1.6& 4. \\
288461 & 0.004396 & 7$a$ & 2. & 53.8 & 223.2!0.1 & 0.61 (149)& 56 (1)& 1.2 & 2.1, 1.3& 0.3 \\
296685 & 0.025415 & 3$a$ & 0.5 & 31.8 & 126.8!1.0 & 0.21 (101)& 32 (2)& 0.9 & 0.4, 1.7& 2.7 \\
296829 & 0.053763 & 2$\star$ & 5. & 63.9 & 139.5!0.3 & 1.08 (89)& 167 (3)& 2.1 & 2.1, 1.9& 15.8 \\
296847 & 0.02563 & 4$\times$ & 0.5 & 63.9 & 108.5!0.1 & 0.34 (144)& 86 (2)& 1.4 & 4.7, 1.7& 2.1 \\
297633 & 0.055052 & 4 & 1. & 41.4 & 353.4!0.1 & 1.3 (152)& 184 (1)& 1.3 & 2.2, 1.7& 40.3 \\
300477 & 0.029288 & 7$\star$ & 1. & 41.4 & 289.7!0.1 & 0.95 (139)& 100 (2)& 1.4 & 3.7, 1.6& 2.9 \\
301346 & 0.04416 & 9$\star$ & 3. & 57.3 & 71.5!0.1 & 2. (117)& 187 (2)& 2.2 & 2.1, 1.7& 18.2 \\
301382 & 0.058449 & 11$\star$ & 5. & 67.7 & 306.5!0.1 & 1.35 (82)& 108 (2)& 1.9 & 1.2, 1.7& 17.4 \\
301799 & 0.051461 & 2$\star$ & 3. & 67. & 321.2!0.7 & 1.13 (51)& 177 (3)& 2. & 3.7, 1.9& 7.3 \\
303099 & 0.026025 & 1$\star$ & 0.5 & 28.4 & 94.1!0.2 & 1.18 (194)& 337 (1)& 1.8 & 6.4, 1.9& 17.1 \\
318936 & 0.017801 & 9$\star$ & 1. & 50.9 & 209.3!0.1 & 1.32 (123)& 66 (2)& 1.1 & 2.9, 1.4& 1.4 \\
319018 & 0.049157 & 5$\star$ & 2. & 60. & 131.4!0.6 & 2.07 (96)& 120 (3)& 1.5 & 0.5$\ddagger$, 1.7& 15.2 \\
319057 & 0.054807 & 8$\star$ & 7. & 65.2 & 258.9!0.1 & 1.11 (155)& 196 (3)& 2.3 & 3., 1.7& 19.0 \\
322910 & 0.031065 & 5$a$ & 0.5 & 35.9 & 279.4!0.8 & 0.47 (162)& 36 (2)& 1.4 & 0.1, 1.7& 7.3 \\
323507 & 0.039919 & 7$\star$ & 0.5 & 30.7 & 37.5!0.1 & 0.84 (130)& 155 (3)& 2.2 & 3.4$\ddagger$, 1.6& 4.4 \\
323855 & 0.040959 & 3$\star$ & 2. & 37.8 & 261.7!0.1 & 1.7 (187)& 209 (1)& 1.3 & 2.7, 1.8& 42.5 \\
323874 & 0.058114 & 5$\times$ & 0.5 & 23.1 & 307.1!0.3 & 0.44 (100)& 92 (3)& 1.7 & 0.3, 1.5& 34.0 \\
325378 & 0.084624 & 1$\star$ & 0.5 & 53.1 & 117.6!1.0 & 0.75 (65)& 259 (2)& 1.5 & 1.4$\ddagger$, 1.9& 74.5 \\
345682 & 0.025492 & 3$\star$ & 1. & 62.6 & 291.2!0.2 & 1.62 (95)& 84 (3)& 1.8 & 1.4$\ddagger$, 1.6& 2.7 \\
346793 & 0.056261 & 3$a$ & 2. & 65.2 & 69.3!0.1 & 3.75 (121)& 186 (1)& 1.4 & 3.5, 1.7& 26.3 \\
346892 & 0.058541 & 3$\dagger$$a$ & 1. & 33.9 & 215.2!0.1 & 1.63 (166)& 171 (2)& 2.1 & 2., 1.7& 24.8 \\
371789 & 0.026657 & 5$\star$ & 2. & 52.4 & 2.6!0.1 & 1.98 (115)& 122 (3)& 2.6 & 1.4, 1.9& 6.7 \\
376001 & 0.051309 & 1$\star$ & 0.5 & 21.6 & 85.6!0.1 & 0.8 (109)& 210 (3)& 2.5 & 2.5, 1.9& 18.9 \\
376121 & 0.051593 & 1$\star$ & 7. & 54.5 & 339.9!0.2 & 1.66 (120)& 287 (1)& 1.2 & 2.1, 1.9& 81.4 \\
376185 & 0.034067 & 4$\times$ & 0.5 & 46.4 & 61.3!0.1 & 1.07 (87)& 71 (3)& 2.1 & 1.2, 1.5& 2.4 \\
376340 & 0.050744 & 3$\star$ & 2. & 63.9 & 199.8!0.1 & 1.45 (97)& 165 (3)& 1.4 & 2.3, 1.7& 23.8 \\
376478 & 0.051329 & 3$\star$ & 1. & 45.6 & 254.3!0.1 & 2.53 (187)& 170 (1)& 1.3 & 3.7, 1.6& 22.2 \\
381215 & 0.050008 & 1 & 0.5 & 70.7 & 139.7!0.4 & 0.82 (71)& 187 (3)& 1.3 & 6.7, 2.& 12.6 \\
381225 & 0.050369 & 3 & 2. & 62.6 & 195.7!0.1 & 1.61 (149)& 180 (2)& 2. & 3., 1.8& 20.7 \\
382152 & 0.056767 & 2 & 0.5 & 43.9 & 61.4!0.1 & 1.29 (191)& 129 (1)& 1.9 & 1.8$\ddagger$, 1.8 & 17.9 \\
382631 & 0.054532 & 3 & 1. & 32.9 & 146.1!0.1 & 1.2 (106)& 167 (3)& 2.5 & 1.9, 1.8& 16.9 \\
383259 & 0.057242 & 1 & 2. & 54.5 & 108.2!0.1 & 1.41 (179)& 98 (3)& 2. & 0.3, 1.8& 40.8 \\
383283 & 0.017288 & 4$\dagger$ & 1. & 68.9 & 206.1!0.1 & 1.05 (78)& 90 (3)& 1.7 & 2.3, 1.8& 1.6 \\
388451 & 0.01218 & 5 & 1. & 70.1 & 47.5!0.2 & 1.19 (117)& 68 (2)& 1.5 & 4.5, 1.5& 0.5 \\
418725 & 0.037778 & 4 & 4. & 28.4 & 65.1!0.1 & 1.48 (150)& 271 (3)& 2.8 & 2.9, 1.8& 23.1 \\
422320 & 0.031514 & 6$\dagger\times$ & 3. & 67.7 & 176.4!0.2 & 1.06 (112)& 106 (1)& 0.9 & 5., 1.5& 5.1 \\
422366 & 0.02867 & 4$\dagger\circ$ & 2. & 59.3 & 168.6!0.2 & 1.04 (144)& 110 (1)& 1. & 2.7, 1.6& 7.9 \\
422619 & 0.028827 & 3$\times$ & 0.5 & 30.7 & 141.3!0.4 & 0.52 (103)& 89 (1)& 0.8 & 3.1, 1.6& 4.7 \\
422761 & 0.076985 & 3 & 0.5 & 29.5 & 256.6!0.1 & 0.5 (168)& 194 (2)& 1.7 & 2.3, 1.8& 44.4 \\
422933 & 0.029037 & 6$\times$ & 2. & 65.8 & 198.3!0.1 & 1.85 (137)& 202 (2)& 1.6 & 2.9, 2.& 14.3 \\
460374 & 0.025131 & 3 & 6. & 46.4 & 25.6!0.1 & 1.23 (184)& 243 (1)& 0.9 & 2.2, 1.9& 60.7 \\
485504 & 0.056051 & 2 & 0.5 & 39.6 & 26.7!0.1 & 0.36 (151)& 187 (2)& 1.4 & 4.1, 1.8& 18.8 \\
485885 & 0.054931 & 3 & 1. & 32.9 & 263.7!0.1 & 1.37 (157)& 169 (1)& 1.5 & 2.9, 1.8& 22.9 \\
486872 & 0.042835 & 3$\circ$ & 3. & 54.5 & 121.2!0.1 & 2.99 (164)& 243 (1)& 1.4 & 2.1, 1.8& 37.6 \\
487027 & 0.026344 & 10 & 8. & 49.5 & 76.5!0.2 & 1.1 (166)& 159 (1)& 1.4 & 2.9, 1.7& 12.1 \\
491552 & 0.024956 & 5 & 0.5 & 65.8 & 89.7!6.3 & 0.67 (144)& 102 (2)& 1.6 & 4., 1.6& 1.9 \\
493621 & 0.029452 & 5 & 1. & 39.6 & 260.2!0.1 & 0.95 (151)& 120 (2)& 1.4 & 5.3$\ddagger$, 1.5& 2.0 \\
504882 & 0.053942 & 2$\star$ & 1. & 35.9 & 169.3!0.1 & 0.63 (104)& 162 (2)& 1.6 & 2.7, 1.8& 16.5 \\
505788 & 0.042837 & 3$\star$ & 7. & 45.6 & 152.3!0.1 & 1.42 (158)& 325 (1)& 1.3 & 3.4, 1.9& 72.5 \\
505817 & 0.043799 & 5$\times$ & 0.5 & 61.3 & 180.6!0.1 & 1.23 (113)& 120 (1)& 1.7 & 8.2, 1.5& 2.5 \\
505979 & 0.043518 & 5$\star$ & 2. & 63.3 & 144.2!0.2 & 0.82 (154)& 156 (2)& 2.1 & 3.5$\ddagger$, 1.6& 8.5 \\
508421 & 0.055285 & 2$a$ & 1. & 42.3 & 11.6!0.1 & 1.68 (138)& 216 (2)& 2. & 3.5, 1.9& 22.7 \\
508481 & 0.056121 & 3$a$ & 1. & 32.9 & 92.9!0.1 & 1.52 (108)& 145 (2)& 1.6 & 2.8, 1.8& 11.6 \\
509444 & 0.03441 & 6$\times$ & 0.5 & 49.5 & 221.5!0.2 & 0.38 (130)& 76 (1)& 1.5 & 5.6, 1.7& 1.2 \\
513066 & 0.029424 & 8$\star$ & 1. & 67. & 353.2!0.2 & 1.55 (117)& 75 (1)& 1. & 3.1, 1.4& 4.3 \\
517070 & 0.050899 & 4$\star$ & 1. & 43.9 & 278.3!0.1 & 2.07 (193)& 159 (1)& 1.8 & 2.6, 1.7& 22.6 \\
517164 & 0.049732 & 1$\star$ & 0.5 & 64.5 & 277.8!0.1 & 1.11 (116)& 191 (2)& 1.3 & 1.6$\ddagger$, 1.9& 29.4 \\
517167 & 0.029883 & 6$\star$ & 1. & 46.4 & 275.8!0.1 & 1.31 (134)& 108 (3)& 2. & 2.2, 1.6& 3.6 \\
517306 & 0.029572 & 7$\star$ & 1. & 50.9 & 250.1!0.3 & 0.95 (163)& 101 (2)& 2.2 & 1.7$\ddagger$, 1.6& 3.9 \\
585359 & 0.035199 & 3$\star$ & 2. & 63.9 & 176.4!0.1 & 1.6 (153)& 137 (2)& 1.6 & 1.6, 1.9& 20.4 \\
585659 & 0.02486 & 4$\times$ & 0.5 & 62.6 & 16.!0.2 & 1.02 (134)& 97 (2)& 1.8 & 4.3, 1.7& 2.0 \\
585755 & 0.040355 & 4$\star$ & 2. & 68.3 & 93.4!0.2 & 1.07 (121)& 145 (2)& 2.1 & 2.8, 1.7& 8.2 \\
592835 & 0.051664 & 5$\star$ & 4. & 63.3 & 295.7!0.1 & 1.92 (142)& 197 (3)& 2.5 & 1.8, 1.8& 34.3 \\
595060 & 0.044346 & 2$\circ$ & 1. & 23.1 & 298.1!0.1 & 1.08 (194)& 205 (3)& 2.1 & 1.9, 1.8& 28.7 \\
599582 & 0.053122 & 2$\star$ & 0.5 & 47.2 & 355.3!0.2 & 0.94 (181)& 99 (1)& 1.1 & 0.6, 1.8& 45.5 \\
599761 & 0.053422 & 1$\star$ & 0.5 & 29.5 & 154.7!0.1 & 1.21 (201)& 302 (1)& 1.3 & 3.9, 1.9& 63.9 \\
600026 & 0.050983 & 6$\star$ & 3. & 43.9 & 302.!0.1 & 1.27 (146)& 212 (1)& 1.7 & 3.2, 1.7& 21.6 \\
610997 & 0.020496 & 6$\star$ & 4. & 38.7 & 301.1!0.3 & 0.82 (155)& 110 (1)& 1.4 & 3.7, 1.7& 2.8 \\
618116 & 0.050937 & 4$\star$ & 2. & 43.1 & 276.4!0.1 & 1.18 (186)& 195 (2)& 1.6 & 3., 1.6& 27.9 \\
618935 & 0.034335 & 6$\star$ & 2. & 66.4 & 118.2!0.1 & 0.96 (130)& 162 (3)& 2.3 & 1.9$\ddagger$, 1.7& 9.0 \\
618992 & 0.05484 & 4 & 10. & 64.5 & 343.7!0.1 & 4.9(59)& 300 (3)& 1.4 & 2.4, 1.9& 63.5 \\
619095 & 0.052577 & 5 & 3. & 47.2 & 270.!0.1 & 1.42 (59)& 320 (3)& 2.8 & 2.2, 1.8& 34.2 \\
619098 & 0.035544 & 5$\dagger$ & 0.5 & 65.8 & 80.5!0.1 & 0.99 (85)& 80 (1)& 0.9 & 3.1, 1.5& 4.9 \\
619105 & 0.025889 & 5$\dagger\times$ & 1. & 63.3 & 214.4!0.1 & 1.13 (151)& 97 (1)& 0.8 & 1.3$\ddagger$, 1.6& 10.4 \\
620098 & 0.027918 & 7 & 1. & 66.4 & 180.3!0.1 & 1.32 (85)& 95 (1)& 0.9 & 3.$\ddagger$, 1.4& 2.2 \\
622429 & 0.040968 & 4$\circ$ & 4. & 58.7 & 135.1!0.1 & 1.71 (136)& 233 (2)& 2.1 & 2.2, 2.& 30.5 \\
622434 & 0.041257 & 2$\times$ & 4. & 51.7 & 182.!0.3 & 2.14 (183)& 264 (1)& 1.5 & 3.1, 1.9& 36.2 \\
622534 & 0.04136 & 4 & 0.5 & 68.9 & 299.4!0.1 & 0.31 (95)& 128 (2)& 2.5 & 3.8$\ddagger$, 1.6& 2.4 \\
622694 & 0.05245 & 3 & 3. & 38.7 & 157.8!0.1 & 1.17 (157)& 205 (2)& 1.6 & 1.2, 1.9& 62.3 \\
622744 & 0.013469 & 11$\dagger$ & 2. & 57.3 & 240.6!0.2 & 0.94 (173)& 77 (2)& 1.9 & 1.2$\ddagger$, 1.5& 1.9 \\
623432 & 0.037998 & 5$\circ$ & 4. & 34.9 & 96.7!0.1 & 1.49 (155)& 302 (1)& 0.9 & 3.6$\ddagger$, 1.8& 39.0 \\
\end{supertabular}
\begin{table*}
\label{tab:table2}
\begin{tabular}{rP.ppcl}
\multicolumn{7}{c}{\textbf{Table 2}} \\
\multicolumn{7}{c}{\diskfit Results for Some of the SGS Cluster Sub-sample\label{tab:fitresults1}} \\
\toprule
GAMA ID & \multicolumn{1}{c}{$z$ Fitted} &   \multicolumn{1}{c}{$\Delta$ISM} &   \multicolumn{1}{c}{Inclination} &   \multicolumn{1}{c}{PA} & $\chi^2$ (DOF) & V$_{\rm max}$ \\
 & &   \multicolumn{1}{c}{(km~s$^{-1}$)} &   \multicolumn{1}{c}{(deg.)} &   \multicolumn{1}{c}{(deg.)} & Fitted & (km~s$^{-1}$)  \\ 
(1) &  \multicolumn{1}{c}{(2)} &   \multicolumn{1}{c}{(3)} &   \multicolumn{1}{c}{(4)} &   \multicolumn{1}{c}{(5)} & (6) & (7) \\ 
\toprule
011327.21+000908.9 & 0.043193 & 2.0 & 45.7!11.7 & -28.2!4.4 & 0.88 (163) & 142 \\ 
011415.78+004555.2 & 0.042379 & 4.0 & 40.1!9.8 & 47.9!1.5 & 0.99 (160) & 173 \\ 
011456.26+000750.4 & 0.041427 & 4.0 & 39.8!11.3 & -75.9!2.3 & 1.26 (165) & 287  \\ 
215604.08-071938.1 & 0.057945 & 4.0 & 38.5!5.7 & 26.7!0.6 & 0.80 (176) & 267  \\ 
215636.04-065225.6 & 0.064572 & 2.0 & 51.8!1.8 & -10.3!0.6 & 0.93 (172) & 197  \\ 
215826.28-072154.0 & 0.060626 & 1.0 & 22.6!5.6 & 58.6!0.7 & 0.77 (145) & 305  \\ 
011346.32+001820.6 & 0.044146 & 4.0 & 68.3!6.9 & -85.1!0.8 & 0.91 (189) & 292  \\ 
215432.20-070924.1 & 0.059227 & 5.0 & 25.1!18.0 & 61.0!2.8 & 1.06 (108) & 511  \\ 
004130.29-091545.8 & 0.044808 & 5.0 & 42.9!5.8 & -1.2!1.6 & 0.94 (154) & 256  \\ 
215705.29-071411.2 & 0.060225 & 1.5 & 35.8!8.6 & -12.5!0.8 & 1.18 (142) & 237 \\ 
215743.17-072347.5 & 0.056766 & 3.0 & 35.2!10.4 & -33.6!2.5 & 1.04 (148) & 332  \\ 
215759.85-072749.5 & 0.05798 & 1.0 & 26.3!7.2 & 27.7!1.5 & 0.82 (137) & 250  \\ 
215853.98-071531.8 & 0.05252 & 5.0 & 58.7!4.8 & -5.3!2.1 & 0.92 (138) & 242  \\ 
215910.35-080431.2 & 0.052561 & 4.0 & 40.9!7.1 & -48.5!1.5 & 1.20 (171) & 261  \\ 
215924.41-073442.7 & 0.058004 & 2.0 & 30.1!8.9 & -39.4!1.5 & 1.23 (143) & 254  \\ 
\bottomrule
\multicolumn{7}{l}{(3) ISM dispersion to normalize \textsc{diskfit} reduced $\chi^2$.} \\
\multicolumn{7}{l}{(5) \textsc{diskfit} evaluated PA is east from N, bootstrapped error.} \\
\multicolumn{7}{l}{(6) Fit (degrees of freedom).} \\
\end{tabular}
\end{table*}
\begin{table*}
\begin{tabular}{rP..pcrl..} 
\multicolumn{10}{c}{\textbf{Table 3}\label{tab:barfitresults}} \\
\multicolumn{10}{c}{\diskfit Results for Some Plausible Weak Bars} \\
\toprule
GAMA & \multicolumn{1}{c}{$z$ Fitted} & \multicolumn{1}{c}{$\Delta$ISM} & \multicolumn{1}{c}{Inclination} & \multicolumn{1}{c}{Disc PA} & $\chi^2$ (DOF) & Bar PA & $V_{\rm t,max}(r/r_{\rm e})$ & \multicolumn{1}{c}{$V_{\rm t2,max}$} & \multicolumn{1}{c}{$V_{\rm r2,max}$} \\
ID & & \multicolumn{1}{c}{(km~s$^{-1}$)} & \multicolumn{1}{c}{(deg.)} & \multicolumn{1}{c}{(deg.)} & Fit & (deg.) & (km~s$^{-1}$) & \multicolumn{1}{c}{(km~s$^{-1}$)} & \multicolumn{1}{c}{(km~s$^{-1}$)} \\ 
(1) & (2) & (3) & (4) & (5) & (6) & (7) & (8) & (9) & (10) \\
\toprule
79635 & 0.04033 &  6 & 53.1 & -42.7!0.9 &   1.0 (192) &  -73 & 209 (1.3) & 11 & 7 \\ 
209807 & 0.05386 &  10 & 41.4 & -187.5!1.4 &   1.1 (177) & -164 & 216 (3) & 41 & 30 \\ 
279818 & 0.02724 &  3 & 37.8 & 197.5!1.9 &   1.2 (177) &  164 & 35 ($>1$) & 3 & 3 \\ 
319018 & 0.04897 &  10 & 60.0 & 136.2!4.8 &   1.7 (120) &  190 & 130 (2) & 16 & 24 \\ 
376121 & 0.05166 &  10 & 54.5 & -19.4!1.3 &   1.9 ( 72) &   -7 & 282 (6) & 23 & 51 \\ 
383259 & 0.05715 &  5 & 54.5 & 110.1!2.0 &   1.1 (146) &  170 & 77 (1.1) & 10 & 19 \\ 
485885 & 0.05492 &  4 & 32.2 & 261.4!0.6 &   0.9 (144) & 278 & 168 (1.1) & 27 & 20 \\ 
508421 & 0.05524 &  10 & 42.3 & 11.2!1.1 &   2.8 (136) &  347 & 225 (1.5) & 13 & 10 \\ 
517070 & 0.05086 &  5 & 43.9 & -80.6!0.8 &   1.1 (178) &   -61 & 164 (1.5) & 13 & 10 \\ 
595060 & 0.04432 &  2 & 23.1 & 295.2!0.9 &   1.2 (189) & 241 & 204 (2.4) & 20 & 35 \\ 
599582 & 0.05307 & 4 & 47.2 & -6.3!2.9 &   1.0 (158) &  69 & 101 (1) & 17 & 19 \\ 
599761 & 0.05333 &  6 & 29.5 & -205.7!0.3 &   1.0 (194) & -150 & 302 (1.1) & 22 & 31 \\ 
619095 & 0.05250 & 10 & 47.2 & -88.2!0.6 &   1.0 (167) &  -146 & 234 (2) & 29 & 11 \\ 
\bottomrule
\multicolumn{10}{l}{(8) CSC maximum velocity in SAMI aperture at the $r/r_e$ shown.} \\
\multicolumn{10}{l}{(9) Tangential $m=2$ motion.} \\
\multicolumn{10}{l}{(10) Radial $m=2$ motion.} \\
\end{tabular}
\end{table*}
\end{center}

\twocolumn

\subsection{Bar Photometric Asymmetries}
The VIKING \citep[VISTA Kilo-degree Infrared Galaxy survey,][]{Edge05} $Z$-band image (Fig.~\ref{fig:ZbandDeconv}) sometimes motivated a dual-\sersic\ decomposition (R. Lange \etal, in prep.): a pair of quite different $n$, PAs, and $r_{\rm e}$.

We interpolated the ``disc'' inward and subtracted it to isolate bulge/bar starlight that we then parametrised with the second \sersic\ profile.
Because central features are compact (Fig.~1), the resulting ``discs'' have mass distributions close to those derived from the single-\sersic\ fits.
We deprojected each image to face-on with the \galfit orientation, rebinned the result into five 1 arcsec wide radial bins and thirty-two 11\fdg25 wide angular bins, then Fourier transformed each angular set and formed the power spectrum up to $m=4$.
Unfortunately, compact features prevented quantification of bar strength by Gaussian fitting the radial extent of their modes \citep{Buta06,Buta07}.

\subsection{Spatially Integrated H~I Velocity Profiles}
H~I content of rotationally supported SGS galaxies came from either the Arecibo Legacy Fast ALFA \citep[ALFALFA,][]{Giovanelli05} survey or early observations from a dedicated follow-up of SGS targets, as described below.

ALFALFA is an H~I-blind survey using the ALFA multi-beam receiver at Arecibo to cover 7000 degrees squared of sky at resolution 3\farcm5 and 5 \kms, and should detect $>30000$ galaxies to $z\la0.06$.
Arecibo observations are limited to -1\arcdeg\ $<\delta<$ 37\arcdeg, so ALFALFA covers only part of the equatorial GAMA fields. 
Furthermore, the current ALFALFA data release \citep{Haynes11} is limited to declinations $>4$\arcdeg, so the H~I spectra presented in \S3.2 were kindly provided by the ALFALFA team before publication.
H~I masses will be published by the ALFALFA team, but our detections were
substantially smaller than $L_{\rm r} \times \Upsilon_\star$ discussed in \S4.3.
Molecular mass should be even smaller.

An ongoing Arecibo program targets SGS galaxies in equatorial GAMA fields that ALFALFA did not detect and uses the $L$-band wide receiver plus same correlator setup as the \textit{GALEX} Arecibo SDSS Survey \citep[GASS,][]{Catinella10, Catinella13}.
Spectra are processed with GASS software, and velocity widths are measured at the half height of each peak using the technique adopted by GASS and ALFALFA.
These deeper spectra were available for six of our sample galaxies.
Unfortunately, radio-frequency interference from the San Juan International Airport radar is preventing study of many SGS discs near $z\sim0.05$.

\begin{figure}
\centering
\includegraphics[scale=0.37]{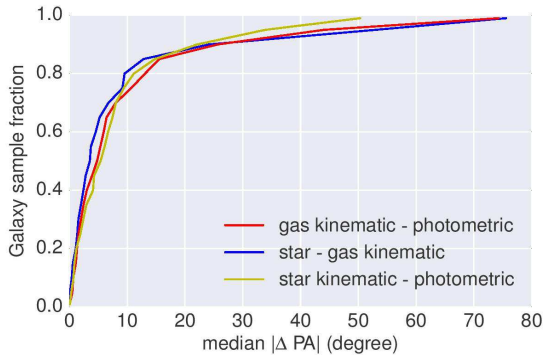}
\caption{\label{fig:angleSpread}Distributions of absolute differences between photometric and the two kinematically established PAs for our GAMA sub-sample, the median values within the SAMI aperture.
To fit, 
\textsc{ppxf} was used for stars, \textsc{diskfit} for gas, and \textsc{sigma}
for photometry.}
\end{figure}

\begin{figure}
\centering
\includegraphics[scale=0.37]{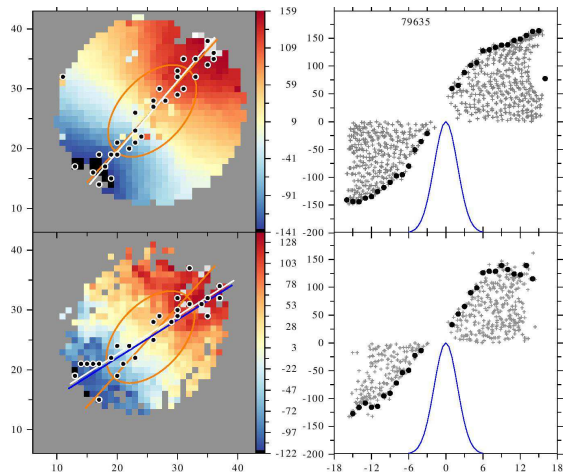}
\protect\caption{\label{fig:gasvsstars}Example gaseous (top) vs.\ stellar CSC from SAMI \textsc{lzifu} fits. 
The pressure supported part of the stellar field is not shown. 
The kinematical line of nodes is traced by the maximum LOS velocity within each elliptical annulus; all points in the cube are plotted at right within the envelope in bold. 
The major axis PSF is shown. 
Ellipse centres come from \textsc{lzifu} fits to emission lines and from \textsc{ppxf} template fits to stellar absorption.}
\end{figure} 
\begin{figure}
\centering
\includegraphics[scale=0.25]{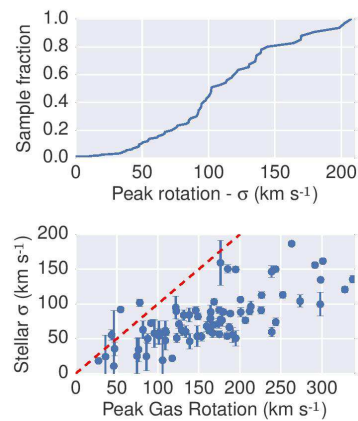}
\protect\caption{\label{fig:rotVSsig}For some of our GAMA sub-sample averaged over the $r_{\rm e}$ ellipse, these plot the distribution in \kms\ of (top) gas peak CSC minus stellar $\sigma$, and
(bottom) gas peak CSC derived by \diskfit vs.\ $\sigma$ from \textsc{ppxf} fits to stellar absorption.}
\end{figure}

\begin{figure*}
\centering
\includegraphics[width=0.74\textwidth,trim=2 2 2 2,clip]{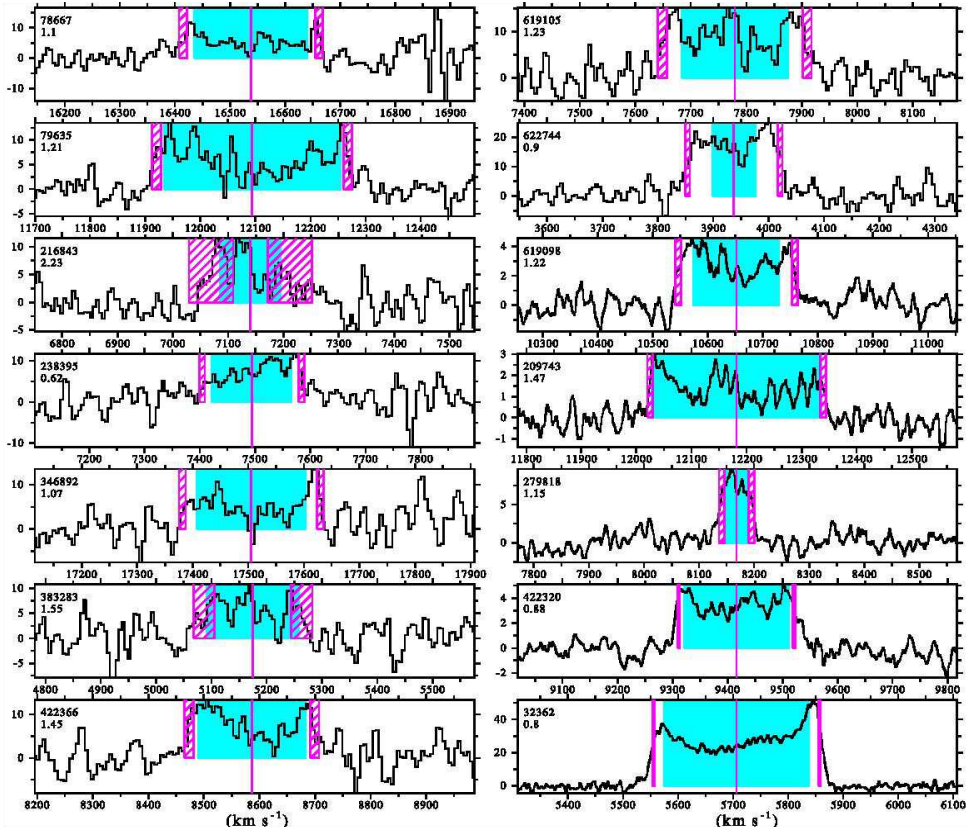}
\protect\caption{\label{fig:HI}
Representative ALFALFA survey H I profiles of SGS galaxies plus 5 longer Arecibo integrations at bottom right.
Horizontal axis is heliocentric velocity (\kms), vertical axis is flux density (mJy) integrated over the 3\farcm5 diameter beam. 
Cyan shades the velocity range of \textsc{diskfit}s to flat discs of ionised gas. 
Magenta shades the W50 $2\sigma$ extents of H~I after correcting for 5 \kms\ FWHM spectral resolution.
The number below the ID gives the asymmetry of flux integrated below/above systemic velocity.}
\end{figure*}

\section{Empirical Results}

\subsection{Stellar/Gaseous Velocities and PAs Compared}
To assess disc orbit regularity and flatness within the SAMI aperture, following \citet{Verheijen01} we plotted the kinematical line of nodes (LON) for both rotational stellar and gaseous velocities, then traced maximum velocity found within each elliptical ring. 
A flat, axisymmetric disc has straight photometric LON. 
A disc with slightly elliptical stellar orbits will show systematic differences in PA between photometric and kinematic LONs where orbits crowd.
Fig.~\ref{fig:angleSpread} summarises such distortions across our GAMA sub-sample,
plotting cumulative distributions of angular differences between gaseous, stellar, and photometrically derived PAs; Fig.~\ref{fig:gasvsstars} shows an example.
Stellar vs.\ photometric values deviate by $>25$\arcdeg\ for 10 percent.
Fig.~\ref{fig:rotVSsig} compares the central stellar LOSVD to the peak gas CSC.
Within a substantial bulge, asymmetric drift will reduce LOSVD below the CSC. 
Almost all trend this way despite small bulges: the CSC peak almost always exceeded stellar $\sigma$ averaged over $r_{\rm e}$ by $\sim120$ \kms.

\subsection{Circular Speed Curves\label{sec:HI}}
Tables 1 and 2 include the observed peak value $v_{\rm max}$ at radius $r_{\rm max}$ from our CSC fits.
Fig.~\ref{fig:HI} compares the peak value of our CSCs with the H~I velocity extent W50/2+$2\sigma$; almost all CSCs attain at least 90 percent of the H~I amplitude.
It also notes the asymmetry of the H~I profile, measured as the ratio of integrated blue-/red-shifted flux.
Despite $\pm20$\ percent variations in this, we found no trend in CSC residuals that would arise from $m=1$ mass asymmetries.
Fig.~\ref{fig:fittingresults1} shows the CSC and $2\sigma$ uncertainty spread at each radius.
Some beam-smear corrected CSCs rose gradually across the SAMI radius, maximising near the \sersic\ $r_{\rm e}$ in Table 1 column (9).
We discuss these fits in \S4.3.

\begin{figure*}
\includegraphics[width=\textwidth,height=\textheight]{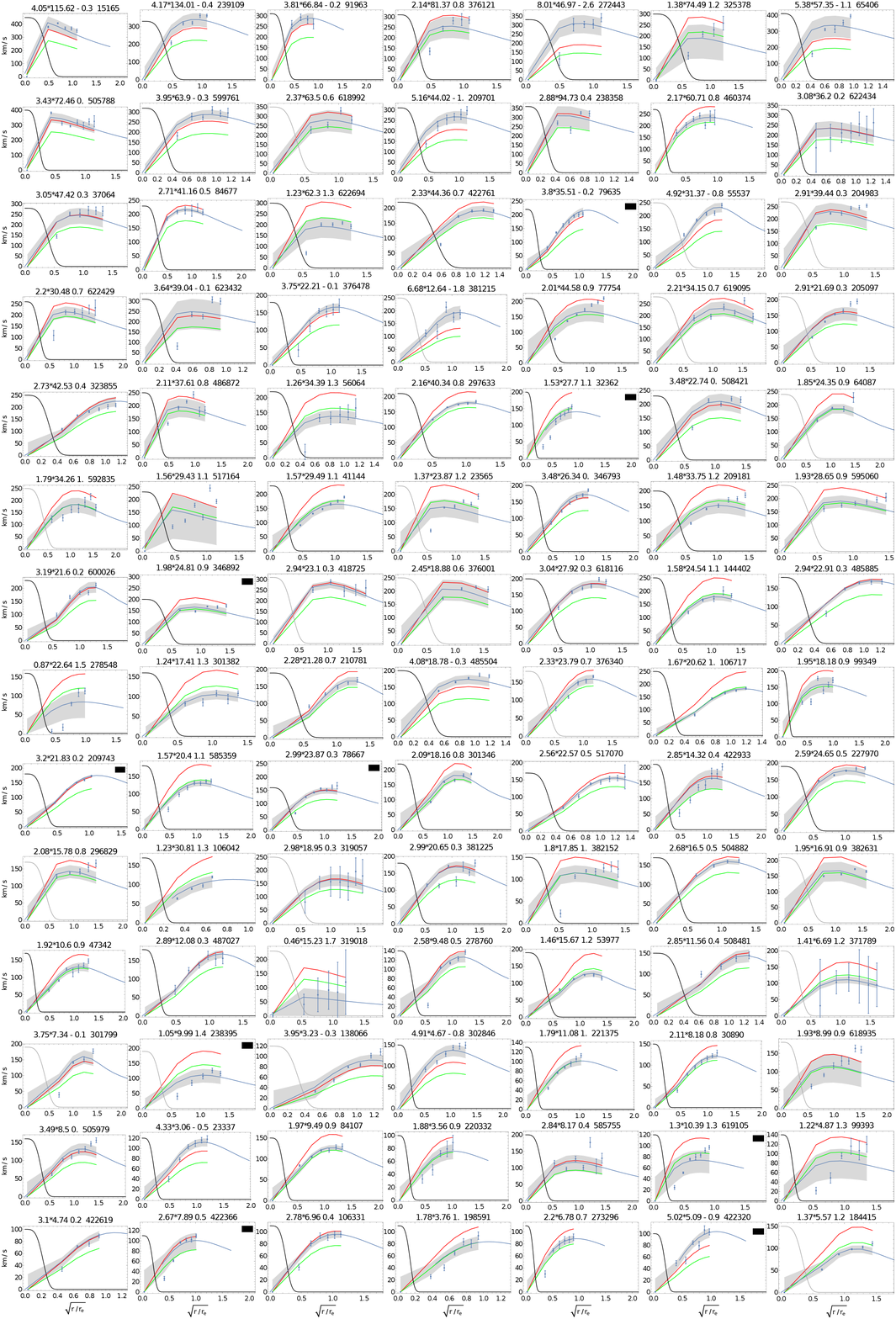}
\end{figure*}
\begin{figure*}
\includegraphics[width=\textwidth,height=0.75\textheight]{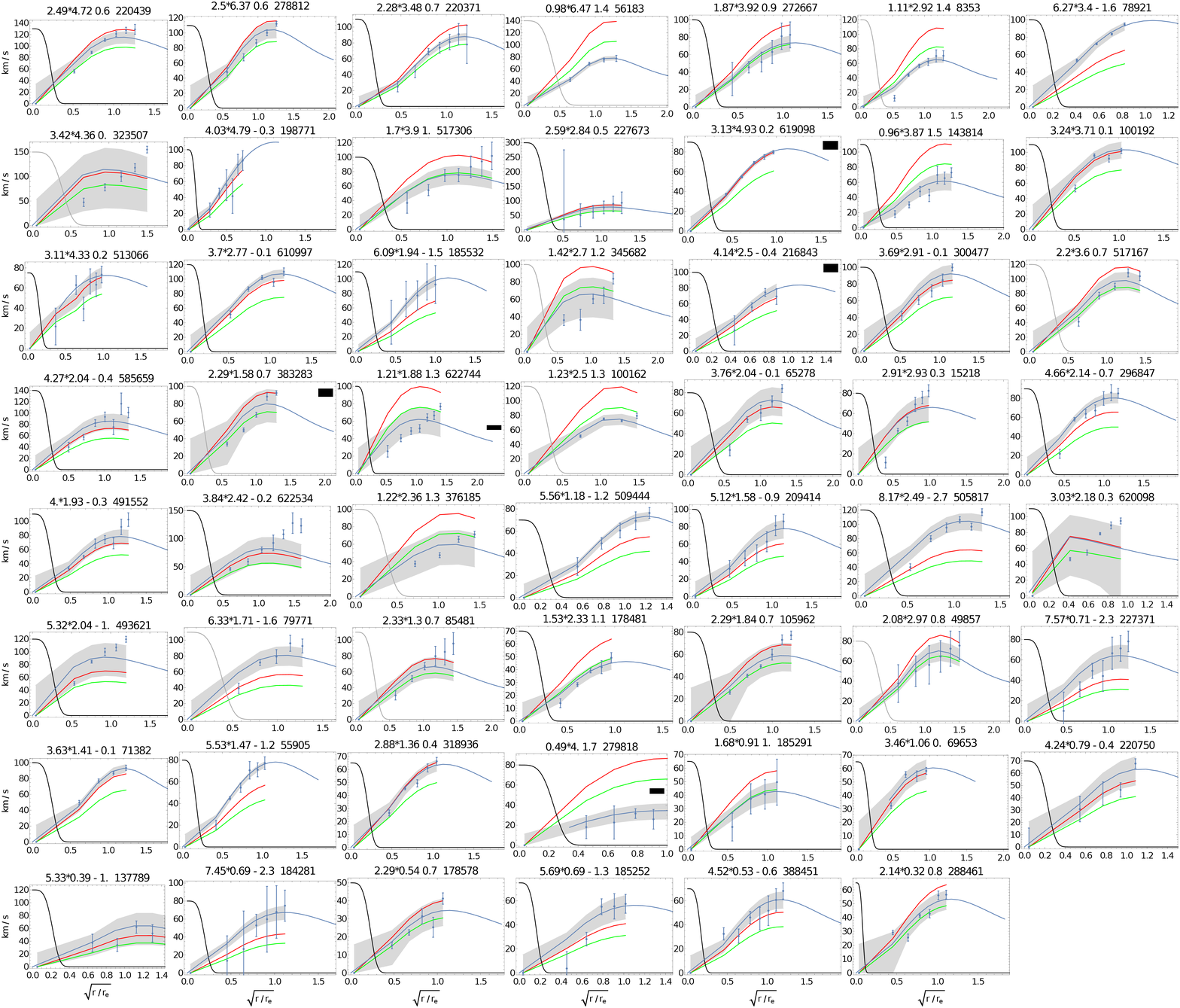}
\protect\caption{\label{fig:fittingresults1}CSCs of our GAMA sub-sample, ordered from maximum to minimum
photometrically derived stellar mass; horizontal axis $\sqrt{r/ r_{\rm e}}$ emphasises small radii.  
The PSF is shown black for sharpest half and gray for the rest.
Uncertainties at each radius span 5--95 percent confidence.
Gray shades the confidence interval of the error-weighted linear fit to the CSC by the $r$-band single-\sersic\ starlight profile of intrinsic flattening $q=0.1$.
The best-fit blue line is extrapolated to the maximum radius of the successful \sersic\ model to show the wedge that opens for DM plus perhaps some H~I to flatten the CSC beyond.
The black box on 14 CSCs spans vertically the uncertainty of the maximum rest-frame velocity of H~I (Fig.~\ref{fig:HI}).
Atop each panel is listed the best-fit $\Upsilon_r$ $\times$ dereddened stellar luminosity in units $10^9M_\odot$, \citet{Kennicutt94} historical star-formation parameter $b$ for the best-fit Salpeter IMF, and the GAMA ID.
$b >1$ denotes an ongoing starburst, and $b<0$ denotes a fit requiring DM.
Red and green curves are the maximum amplitude CSCs generated by Salpeter and Kroupa IMFs for $b=0$, respectively.
When they lie below the shaded gray band, DM is required within the SAMI aperture.
}
\end{figure*}

\begin{figure*}
\centering
\includegraphics[width=0.8\textwidth]{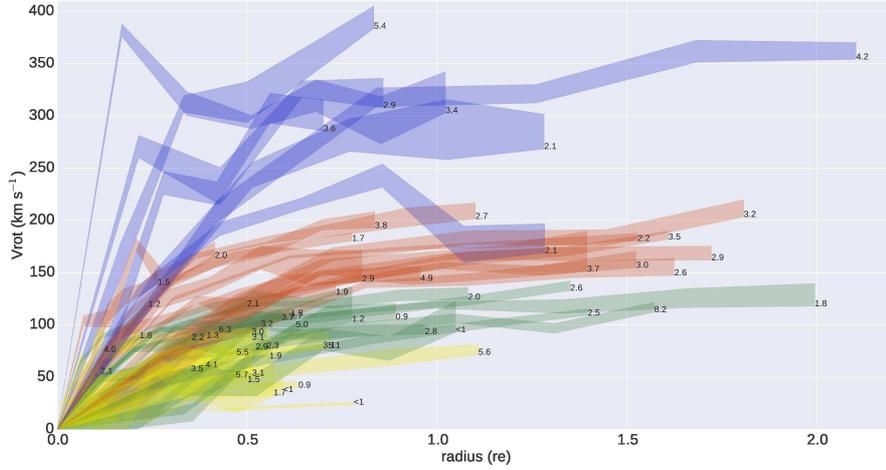}
\protect\caption{\label{fig:rotbyamplitudes}CSCs having median or smaller beam-smear; each is labelled by its median dynamical \MXL in solar units after dereddening and correcting for line emission within the $r$-band. 
Horizontal axis is major-axis radius in \sersic\ $r_{\rm e}$ scale lengths.
Colour shading of the CSC fit 9--95 percent confidence bands merely attempts to reduce confusion.}
\end{figure*} 

\begin{figure}
\centering
\includegraphics[width=0.35\textwidth]{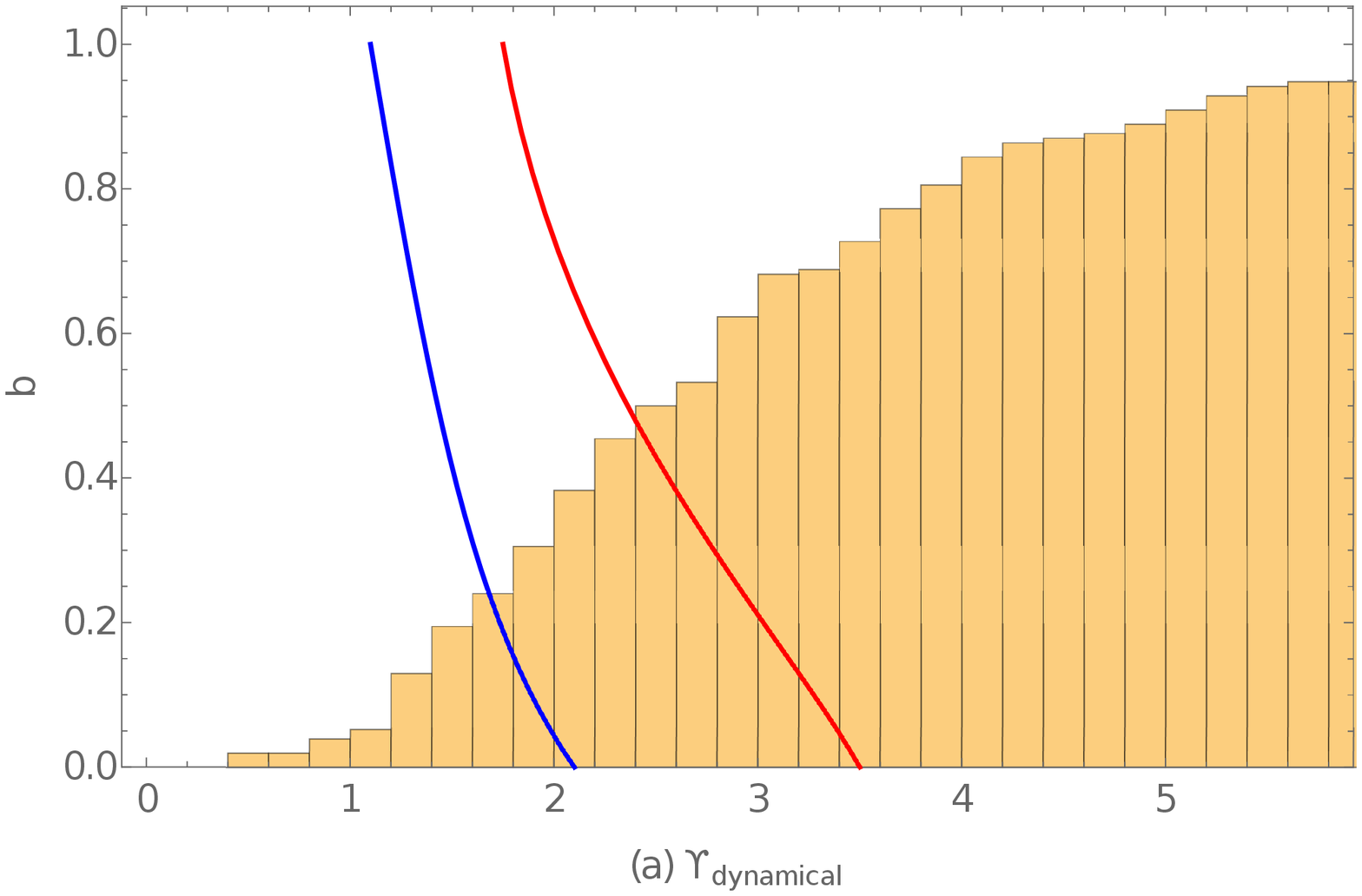}
\includegraphics[width=0.35\textwidth]{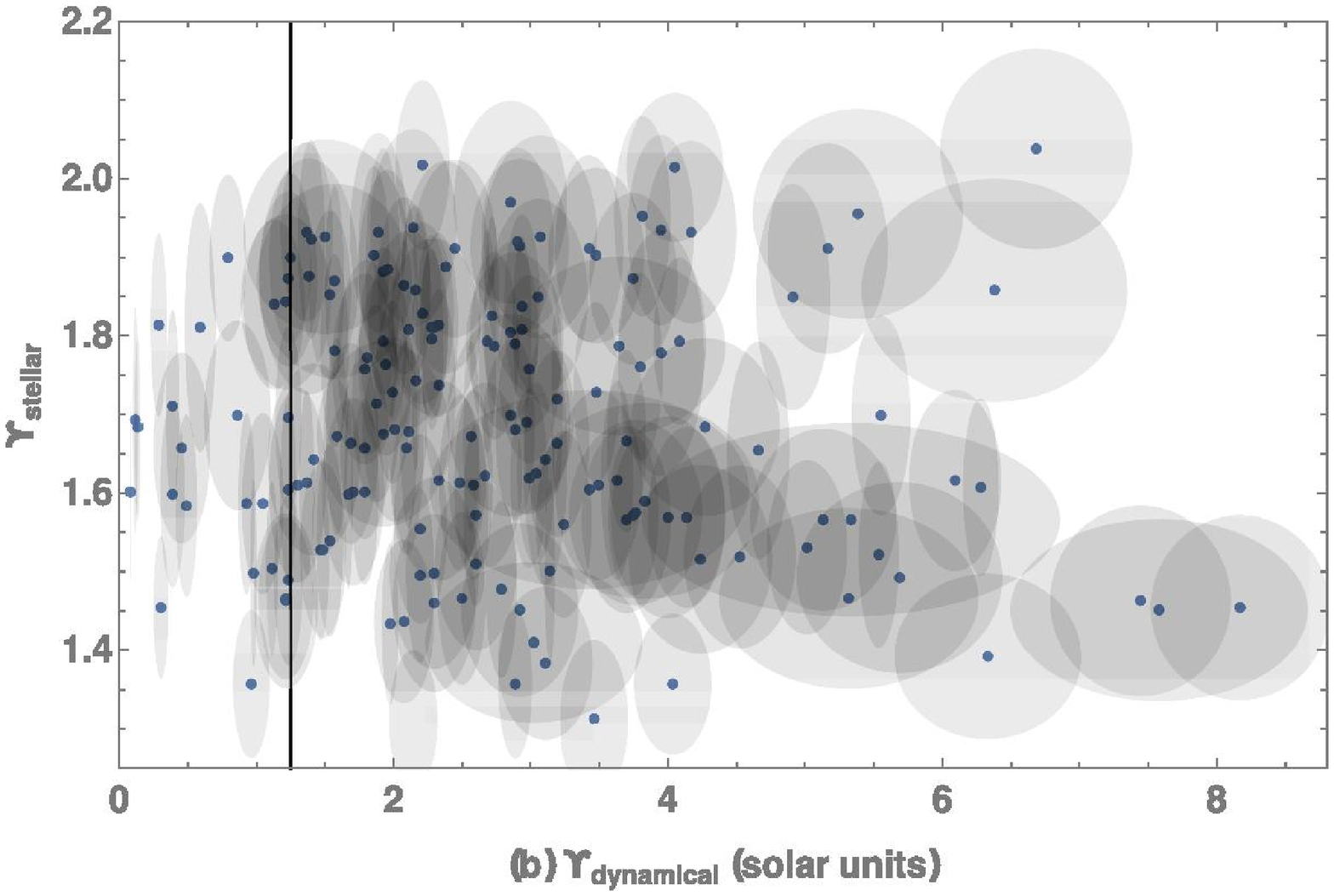}
\protect\caption{\label{fig:MtoL}
GAMA sub-sample (a) cumulative distribution of derived \MXL\ assuming intrinsic flattening $q=0.1$.
Curves and vertical scale indicate the effect of \citet{Kennicutt94} historical star-formation rate (SFR) parameter
$b\equiv$ SFR$_{\rm current}$/$\langle{\rm SFR}\rangle_{\rm past}$.
CML model values to left of the red curve are valid for exponentially declining SFR with two IMFs from PST04 fig.\ 4: red is Salpeter, blue is Kroupa.
The ``diet Salpeter" IMF of \citet{Bell01} falls between these.
(b) IMF-unconstrained \MXL compared to $\Upsilon_{\rm P}$ derived from stellar population fits across the brightest 40 percent of a galaxy; uncertainty ellipses span $2\sigma$.
The two $\Upsilon$ measures are clearly uncorrelated.}
\end{figure}

\subsection{Mass Estimates}\label{sec:massmodels}
Column 
(10) of Table 1 reports $\Upsilon_{\rm P}$ (1.7 median) from fits to the $ugriz$-starlight defined SED of the GAMA sub-sample.
These revealed that the starlight weighted ages are Gaussian distributed with mean 3.1 Gyr and dispersion 0.6 Gyr.
For the GAMA sub-sample, Fig.~\ref{fig:MtoL}a summarises the distribution of median \MXL (2.6 median) and the product of this with the dereddened starlight extrapolated to $10r_{\rm e}$ to give the mass distributed like starlight while ignoring an IMF constraint.
To recover some of the $\Upsilon_\star$ fits that failed (i.e.\ whose $b<0$), we adjusted $\Upsilon$ to posit that unobserved baryons+DM distribute like starlight, namely as a very flattened mass homeoids.
Indeed, this \MXL scheme fitted 37 of 42 failures of our CSCs (1/4), including three of our H~I observed galaxies.
Fig.~\ref{fig:MtoL}b compares \MXL and $\Upsilon_\star$ from our fits.
Green in Fig.~\ref{fig:fittingresults1} shades the 5--95 percent confidence interval of the \MXL fits.
Recall that reducing the intrinsic flattening of the homeoids from $q=0.1$ to 0.05 would reduce \MXL by 0.76, increasing consistency with $\Upsilon_\star$.

Most \MXL fits are good, but for the remaining $\sim10$ percent, our $\Upsilon_\star$ fit even using the Salpeter IMF exceeds \MXL, an unphysical result, or the observed CSC deviated strongly at small and/or large radii from the scaled light CSC.
Evidently in these systems, mass does not follow starlight in the SAMI aperture.

\subsection{Streaming}
Even binned, many of our spectral cubes could not map reliably the $h_3$ and $h_4$ moments of the Gauss-Hermite parametrisation of the stellar LOSVD, so we considered only gas streaming.
Our $\ga2$ arcsec resolution prevented study of nuclear bars and rings, so we sought substantial non-axisymmetric features at least 3 arcsec long, $\sim5$ kpc.
The small areal coverage of SAMI compared to e.g.\ CALIFA prevented determination of bar pattern-speed to locate resonances in the differentially rotating disc \citep{Tremaine84}.
To maximise the number of points to compare the axisymmetric with non-axisymmetric fits, we considered the beam-smeared cubes; our fits always allowed the bar major-axis PA to vary and for the kinematical centre to shift by up to $\pm$0.5 arcsec from nominal.
We found that some photometric/kinematic fits co-aligned in PA, but $\chi^2$ was often flat over 10s of degrees in bar PA.
That indeterminancy likely arose because some photometric asymmetries twisted in PA across the SAMI aperture -- for example, a compact bar ending with strong spiral arms -- but tracking a twist was impractical with \diskfit.
For an acceptable non-axisymmetric fit, we required same-sign amplitudes for both
the radial and tangential components of $m=2$ streaming over $>80$ percent of the fit uncertainty range.

\begin{figure*}
\centering
\includegraphics[width=0.98\textwidth]{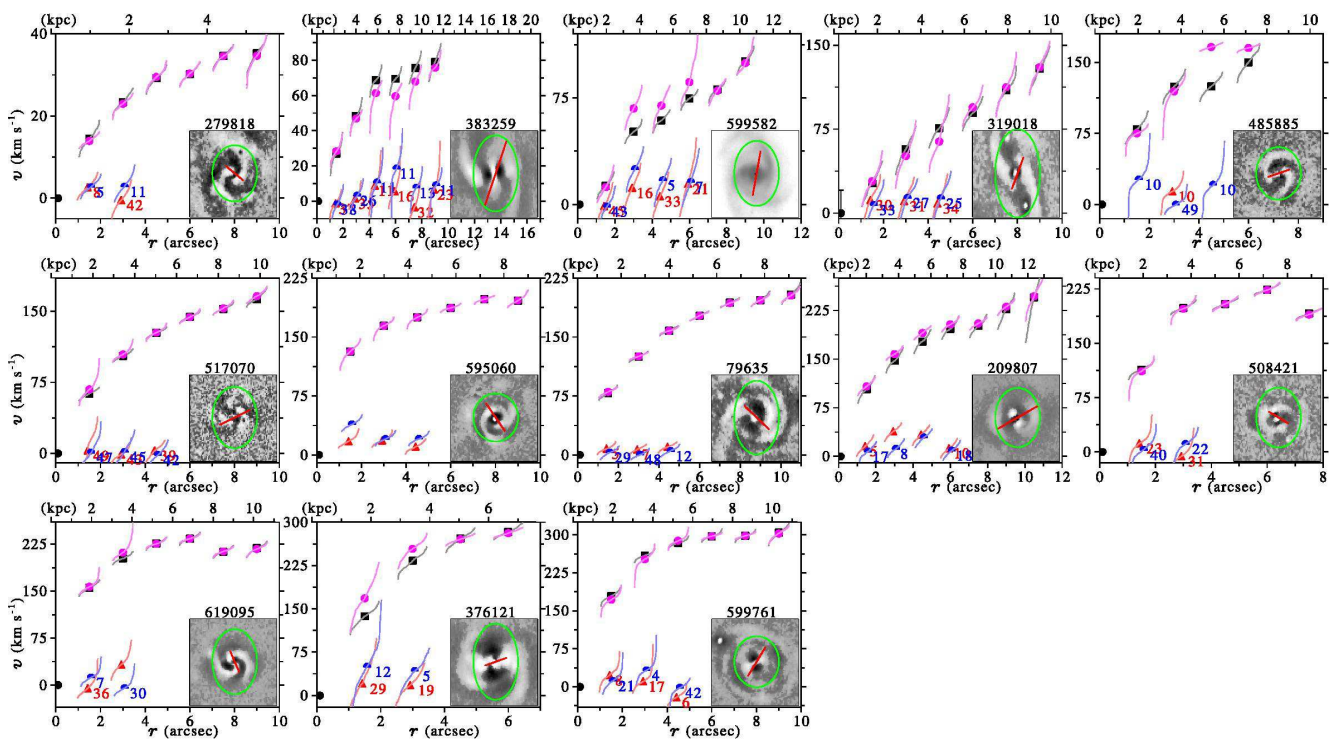}
\protect\caption{\label{fig:barfits}CSCs with (magenta) and without (black) streaming for systems with asymmetric VIKING $Z$-band residuals, ordered by increasing peak circular speed from top left to lower right. 
Inset figure has sqrt-scaled intensities after subtracting the best-fit dual-\sersic\ profile (see Fig.~\ref{fig:streamlines} for detail); 
599582 as yet has no \sersic\ $Z$-band fit so is shown unmodified.
Images are oriented with kinematical major axis horizontal and are deprojected to face-on.
SAMI spanned the green ellipse.
The extent and orientation of the best-fit kinematical distortion is shown by a red line.
Each symbol is plotted at the median velocity of the cumulative uncertainty distribution at that radius, shown from 5 to 95 percentiles. 
Numbers on these distributions indicate the percentage below the 0 \kms (i.e.\ $v_{\rm sys}$) line. 
Black squares plot the axisymmetric CSC, blue semi-circles plot radial streaming, red triangles plot tangential streaming, and magenta circles plot the barred CSC; barred points are displaced slightly for clarity. 
Horizontal axis has arc-second scale at bottom and kpc at top.
}
\end{figure*}

\begin{figure*}
\begin{centering}
\includegraphics[width=\textwidth]{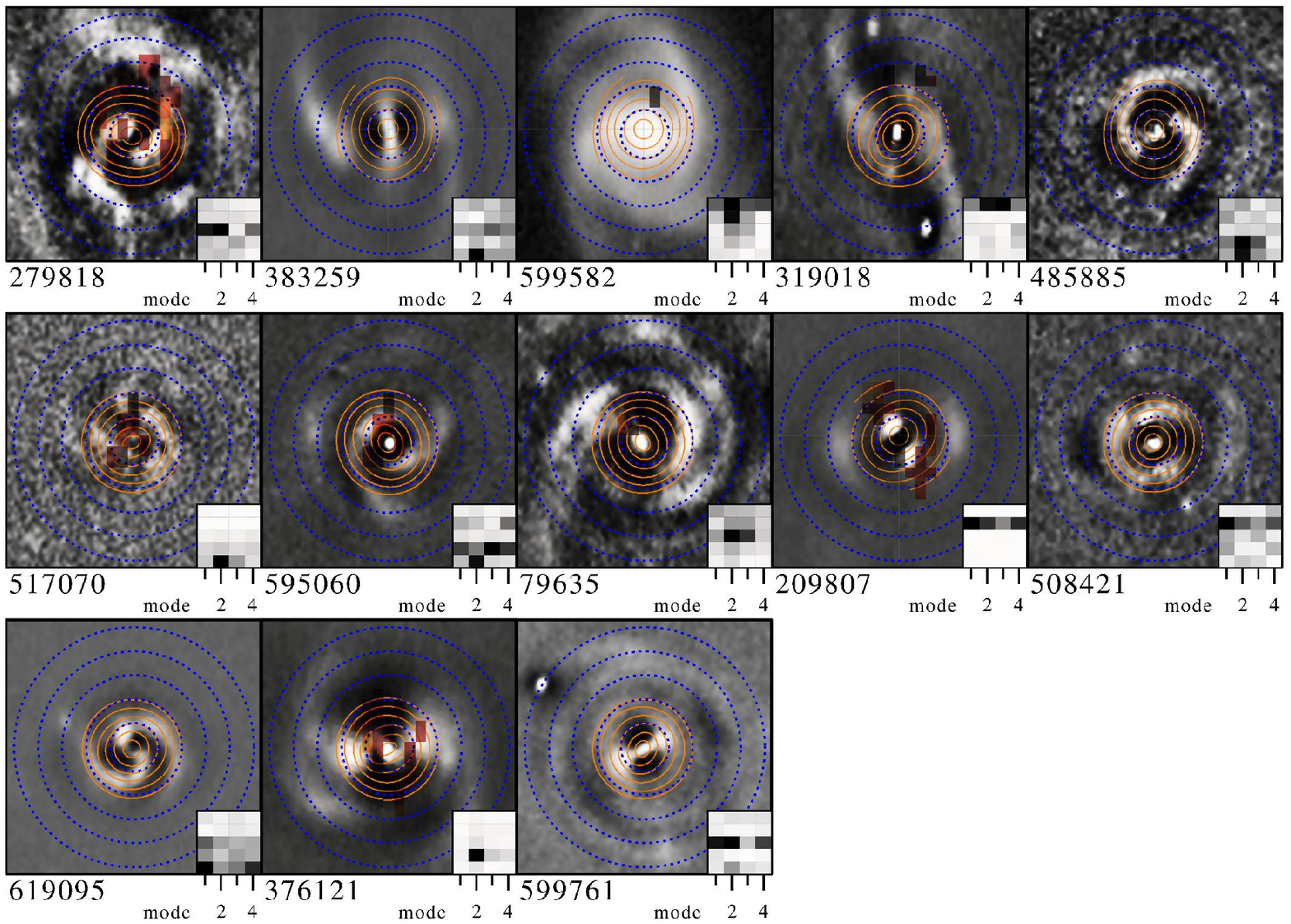}
\protect\caption{\label{fig:streamlines}$Z$-band residuals after subtracting the best-fit dual-\sersic\ profile, deprojected to face-on using the inclination from the \sersic\ single-profile fit, plotted with the kinematic major axis horizontal (37 arcsec extent), and showing in orange the streamlines of our best-fit bar model gaseous velocity field over the SAMI extent left to right. 
599582 as yet has no \sersic\ $Z$-band fit so is shown unmodified.
The inserts show modal content in each blue ring of 3.5 arcsec extent from centre at bottom; note sometimes strong $m=2$ mode from bar (second radial bin) and arms (beyond).
Coloured rectangles on several systems locate high-velocity (150 to 300 \kms from lighter to darker red) shocks as evidenced by diagnostic emission-line flux ratios.
}
\end{centering}
\end{figure*}

Fig.~\ref{fig:barfits} shows non-axisymmetric \textsc{diskfit}s to 13 photometric asymmetries -- bar/oval candidates.
Fig.~\ref{fig:streamlines} shows their angular modes and slightly non-circular streamlines.
The Fourier modal analysis detected such motions down to a few \kms\ in the otherwise most regular discs.
Only ID 595060 and 376121 have in-plane streaming of $>35$ \kms\ but still far below
the shock velocities indicated by their emission-line ratios overlaid in Fig.~\ref{fig:streamlines}.
In several cases discussed in the Appendix, streaming was more plausibly associated with strong spiral arms at larger radii than along a bar.

\section{Discussion}
A flattened bulge may merge into the disc if it results from, or is substantially modified by, a bar where some stars and gas have non-circular orbits.
$\Upsilon_\star$ may vary with radius.
Gas turbulence can make a disc spatio-kinematically ``lumpy''.
We now assess these issues before considering DM.

\subsection{Photometric vs.\ Gas/Stellar Kinematical PAs}
Many galaxies show bars, especially in the near-infrared \citep[at least 70 percent,][]{Eskridge00}. 
Sharper images reveal more nuclear bars, rings, and lenses with stars often forming in a central light cusp \citep[e.g.][]{Bureau99}.
In 12 nearby discs, \citet{Seidel15} find that strong bars do not alter global rotation, but the majority do have double peaks in stellar-rotational velocity along the kinematical major axis, and occasionally show central decreases in velocity dispersion probably from nuclear rings.
\citet{Emsellem07} probe velocity field irregularities and find a dip at $\sim$1/4 bar length in the majority of their sample regardless of Hubble type and bar strength; this is the scale of the Inner Lindblad Resonance \citep{Pfenniger90} that often coincides with a stellar ring or lens \citep[e.g.][and references therein]{Buta12}, but was often within the PSF of our bar candidates so could not be assessed.

\citet{Fred14} (BB14 hereafter) find $<~14$\arcdeg\ differences between stellar and gaseous kinematical PAs for almost all of their barred sample from the CALIFA survey \citep{Sanchez12}.
Our Fig.~\ref{fig:angleSpread} blue curve shows that 15 percent of SGS GAMA sub-sample galaxies exceed that angle.
The CALIFA PA fits span 24$\pm5$ arcsec radii (Table~1 BB14), which is comparable to our fit extent because the SGS averages $\sim3\times$ the redshift of CALIFA (Fig.~2c).
BB14 find that 26/27 barred galaxies have kinematical vs.\ photometric major axis PAs misaligned by $<20$\arcdeg, like unbarred systems; the single large misalignment is attributed to external interaction.
Our Fig.~\ref{fig:angleSpread} red and green curves tested their conclusion.
With more than triple their targets, we found that 5 percent of our sample misalign by $>30$\arcdeg yet show no evidence for a recent external influence.

Bar response of gas should exceed that of stars, e.g.\ strong bars in \citet{Seidel15} have $\sim2.5\times$ the torque in gas than in stars.
\citet{Epinat08} found 15 percent of spirals misaligned by $>15$\arcdeg, our
Fig.~\ref{fig:angleSpread} confirms this fraction.

In summary, deviated PAs are evident in the SGS, so we now consider how those flows influence the CSCs.

\subsection{Weak Influence on CSCs from Common Bars}
Strong stellar bars can deviate gas by 30 to more than 100~\kms\ from circular rotation along their leading edge \citep[e.g.][using \diskfit]{Sellwood10}.
An axisymmetric fit to a distorted velocity field will be biased low when the streaming distortion aligns near the disc major axis because gas there is at its orbit apogalacticon. 
Gas motions along our sightline increase when we view a bar end-on.
It is therefore important to check over broad ranges of bar orientations and strengths that ionised gas moves in sufficiently circular orbits to trace mass density over the full ranges of galaxy environment and mass.
\citet{Holmes15} apply \diskfit to the CALIFA survey DR1 \citep{Sanchez12}, finding non-axisymmetric motions in 12/37 galaxies with 11/12 having bars.
Yet, BB14 and \citet{Seidel15} show that even strong bars do not modify gas CSCs.

We found that the distortion of CSC by a common, weak bar was generally only 10--30 \kms.
The maximum amplitude detected may have shrunk because Figs.~12 and 13 show that the photometric residuals twist in PA over the SAMI aperture, averaging extreme values over the fixed bar axis of our \diskfit\ model.
Indeed, the presence of shock-excited emission-line ratios in more than 2/3 of our galaxies with shock speeds
up to several hundred \kms\ do indicate that such motions are buried in our cubes.
In even our least smeared quartile, $m=2$ distortions were compact, altering only two CSCs 
(376121 and 209807) by $>30$ percent amplitude at small radii as they built up.
Thus, asymmetries in the SGS did not undermine application of a CSC to mass studies, and support the assertion of e.g.\ \citet{Kormendy12} that a locally increasing or flat CSC like all here inhibits secular disc evolution. 

\subsection{Mass Decomposition}
If nominal $\Upsilon_\star$ is increased \textit{ad hoc} to avoid substantial DM (i.e.\ $<25$ percent contribution to CSC) at radii up to peak velocity, i.e.\ our \MXL fits, even compact photometric features can have a kinematical counterpart \citep[e.g.]{Corradi90}.
Of course, a DM halo or MOND must still flatten the CSC in any outer H~I disc beyond most starlight.
In luminous SA galaxies \citet{Noordermeer07} find that the inner CSC shape correlates with the light distribution, so large bulges dominate dynamics.
However, \citet{deBlok08} find correlations even in Sb--c galaxies, 
Likewise, \citet{Salucci08} obtain excellent starlight-only fits for 18 spirals within $3r_{\rm e}$.

In contrast, such ``maximal discs'' have been excluded in 30 nearly face-on galaxies by the DiskMass survey of vertical velocity dispersions \citep{Martinsson13}. 
There, up to half of the disc mass may be DM.
Likewise, the Milky Way Galaxy's CSC can be matched with 40 percent DM within the inner 10 kpc \citep{Portail15}; observations do not exclude some concentration near the disc plane \citep{Bienayme14}.

Lacking H I maps to $\sim20$ kpc \citep{Sofue13}, it was fruitless for us to specify a DM halo form for the quarter of our sample that $\Upsilon_\star$, or the 10 percent that \MXL, failed to fit.
Even with H~I, one can play off luminous mass against a dark halo modified by uncertain compression \citep[e.g.][]{Dutton05}.
For example, \citet{Noordermeer07} isolate discs from large SA bulges to fit CML $\Upsilon_{\rm R}\sim1-10$.
Even with H~I at larger radii, their fits are comparable using various DM distributions.
They find that a spherical, isothermal halo generally fits better especially at $>10$ kpc, but for a subset H~I scaling works too.
\citet{deBlok08} are sensitive to DM halo details because their H~I data are extensive.
With this background, we now discuss our findings on \MXL then $\Upsilon_\star$.

\subsubsection{Trends of Dynamical and Stellar M/L}
As the distribution of luminous mass flattens, $V(r)$ increases while $\Upsilon V^2(r)$ remains constant, so $\Upsilon$ declines.
As extremes, \citet{Takamiya00} bound $\Upsilon_{\rm V}$ by placing all mass in either a sphere or disc.
They find that VML $\Upsilon_{\rm V}$ on average at most doubles to $\sim10$ solar from 2 kpc to our typical maximum extent of 10 kpc. 
This contradicts the chemo-photometric models of PS10, whose disc VML $\Upsilon_\star(r)$ declines at larger radii and from smaller values $\la2.5$. 
Indeed, in all inside-out galaxy formation models, VML from stars declines with increasing radius from younger/less metallic stars there \citep[e.g.][PS10 hereafter]{Portinari10}, but cannot generate the full amplitude of the CSC.

PS10 include a DM halo and fit Sb--Sc CSCs within $2R_{\rm opt}$ equally well with either CML or VML, finding that a VML/CML disc contributes less/more to the CSC hence mass and implies a larger/smaller DM core.
More rapid mass buildup with radius draws the peak of the CSC inward by $\sim25$ percent radius while boosting its amplitude $\sim10$ percent.
The VML stellar mass profile is 20 percent more centrally concentrated than the light.

With VML, PS10 predict an expanded region of ``inner baryon dominance'' and suggest that it may also explain unusually strong radial colour gradients in some galaxies.
They approximate the $I$-band stellar $\Upsilon_\star$ normalized at $r_{\rm e}$ by
\begin{equation}
\frac{\Upsilon_\star(r)}{\Upsilon_{\star,1}} =\exp\bigg\{-\alpha(s)\bigg[\Big(\frac{R}{h_I}\Big)^s-1\bigg]\bigg\}
\end{equation}
with $\alpha(s)=1.25 (1.3-s)^3+0.13$, and $s=1\pm0.1$ for the ``shallow'' form (essentially linear within $3r_{\rm e}$) and $0.6\pm0.1$ for the ``concave'' form (``cuspy'' within $0.5r_{\rm e}$).
The bottom panel of PST04 fig. 6 shows comparable spread between models in $R$-band (almost our $r$-band) but rising faster at small $b$ than in $I$-band.

Small $r_{\rm e}$s meant that the SGS could not distinguish between these forms: uncertainty bands of our \MXL fits often spanned both.
CML with either exponentially declining star formation rate or \citet[][hereafter PST04]{Portinari04} chemo-photometric models both fitted our CSCs more often than did VML.
Hence, we discuss only CML \MXL models, which merely up-scaled our \sersic-fitted CSC.

\subsubsection{Implications of Our CML Fits}
Fig.~\ref{fig:rotbyamplitudes} does not show a trend between the $\Upsilon$ and rotational peak velocity (mass) of the successful fits, e.g.\ lower mass systems do not have larger $\Upsilon_\star$.

Two-thirds of our
\MXL distribution is consistent with CML for stars+remnants using various IMFs (example curves in Fig.~\ref{fig:MtoL}a) in PST04 and exponentially declining SFR.
For example, PST04 fig.\ 5 attains $0.75\la\Upsilon_{\rm I}\la3.4$ at solar or greater metallicity as \citet{Kennicutt94} birthrate parameter $b$ ranges from 1 to 0.
Their table 7 shows $\Upsilon_{\rm I}\sim2.1-0.8$ for Sa-Sc discs, respectively.
$\Upsilon_{\rm I}<0.75$ is for $b>1$, i.e.\ star-bursting systems.
Even the lower-mass-heavy Salpeter IMF cannot exceed $\Upsilon_{\rm r}=3.4$, so galaxies in the top 20 percent of our distribution to right of the red curve require more than the stellar mass in unobserved baryons+DM close to $2r_{\rm e}$.
For the PST04 chemo-photometric VML, all IMF curves shift leftward, increasing needed matter within and somewhat beyond $\sim2r_{\rm e}$.
$\Upsilon_\star$ values from our photometric stellar population fits unphysically exceed the dynamically derived \MXL for only 10 percent of our GAMA sub-sample.
Excluding this group,
the distribution agrees with the Salpeter IMF for all $b$ in the PST04 models (red curve in Fig.~\ref{fig:MtoL}a).

\citet{Kennicutt94} find $b<0.2$ for Sa-Sab discs, 0.3--0.4 for Sb, and 0.8--1 for Sbc-Sc.
Therefore, our two-thirds $\Upsilon_\star$ fit successes need few unobserved baryons and/or DM distributed similarly to starlight within 10 kpc radius.
Our median $\Upsilon_\star$ implied 30--40 percent of the starlight mass in H~I, a global result weighted outside the SAMI aperture, which is consistent with most of the ALFALFA fluxes (Fig.~\ref{fig:HI}).
Fig.~\ref{fig:MtoL}a shows that the few failures of \MXL in Fig.~\ref{fig:fittingresults1} imply different distributions of their mass and starlight.

\section{Conclusions}
We studied \targets\ galaxies in the SGS v0.9 release not close to edge- or face-on and lacking prominent $r$-band bars in SDSS images.
We thus developed procedures to quantify CSCs and $\Upsilon_{\rm r}$ for typical galaxies in the SGS having moderately distorted CSCs hence mostly circular orbits.

We used \diskfit\ to map gas kinematics across the whole disc at once to highlight asymmetries and to trace the CSC.
Fit uncertainties were estimated by boot-strapping over PA uncertainties always, sometimes light centre and, for the ``cluster" sub-sample, bar PA using GAMA survey photometric priors; inclinations were fixed at GAMA priors for that sub-sample.

We quantified $m=2$ gas streaming along photometric asymmetries.
We found statistically significant deviations of $<40$ \kms\ in 12 galaxies that however barely altered the CSC, and 80 \kms\ in one.
In this study, emission-line ratios often indicated shocks of several hundred \kms.
The absence of velocity residuals of comparable amplitude was likely a result of compact shock fronts being blurred away by seeing.
For the rest, adding $m=1$ ``lop-sided'' asymmetry did not improve fits.

Many CSCs rose slowly through the SAMI aperture
even after correcting for beam smearing
within $0.5r_{\rm e}$, to peak at 1--2 $r_{\rm e}$ and 80--300 \kms. 
Amplitudes of 14 with representative H~I velocity profiles almost always matched the peak of the SAMI CSC, so those CSCs do not rise beyond the SAMI aperture and are plausibly flat there.

We assumed that mass across the SAMI aperture was distributed in flattened, luminous nested homeoids, and quantified it dynamically -- modulo CML \MXL\ -- from \sersic\ profile fits to $r$-band SDSS photometry.
After correcting luminosity down for line emission and up for average dust attenuation, two-thirds of the distribution of median \MXL (its median = 2.6) was compatible with plausible IMFs and \citet{Kennicutt94} historical birthrate parameter $b$.
We could fit the CSC of 37 more galaxies simply by up-scaling the starlight profile by a CML \MXL.
Some of those would be compatible with $\Upsilon_\star$ if intrinsic flattening of the
disc was reduced from $q=0.1$ to 0.05.
Thus we inferred for those at most a comparable mass of unobserved baryons and/or DM distributed like starlight.

For the remaining $\sim10$ percent, we needed mass distributed quite differently from starlight; more sophisticated population fits to SGS \textit{spectra} using stellar VML may define the radial variations of this offset to isolate the DM.
Our results demonstrate that the
full SGS will be able to explore environmental effects on CSCs, $\Upsilon$, and IMFs.

\section*{Acknowledgements}
GC thanks JBH and the School of Physics at University of Sydney, and the Moseley Fund at UNC, for sabbatical support, and Australian Astronomical Observatory Director Warrick Couch for hospitality. 
We thank J.\ Sellwood for providing his \diskfit code.
We thank R.\ Giovanelli, M.\ Haynes, and D.\ Obreschkow for access to ALFALFA spectra in advance of publication.
We appreciate the constructive comments of the referee.

The SAMI Galaxy Survey is based on observations made at the Anglo-Australian Telescope. 
The Sydney-AAO Multi-object Integral field spectrograph (SAMI) was developed jointly by the University of Sydney and the Australian Astronomical Observatory. 
The SAMI input catalogue is based on data from the Sloan Digital Sky Survey, the GAMA Survey and the VST ATLAS Survey. 
The SAMI Galaxy Survey is funded by the Australian Research Council (ARC) Centre of Excellence for All-sky Astrophysics (CAASTRO), through project number CE110001020, and other participating institutions. 
Many datacubes used in this paper are available at the SAMI Galaxy Survey website \url{sami-survey.org}; eventually derived CSCs and uncertainties will be available there too, but can be obtained from author GC now.
This paper uses GAMA survey data products (\url{www.gama-survey.org}), in particular the \sersic\ photometry DMU which used \galfit available from \url{users.obs.carnegiescience.edu/peng/work/galfit}
GAMA is a joint European-Australasian project based around a spectroscopic campaign using the Anglo-Australian Telescope. 
The GAMA input catalogue is based on data taken from the Sloan Digital Sky Survey and the UKIRT Infrared Deep Sky Survey. 
Complementary imaging of the GAMA regions is being obtained by numerous independent survey programs including \textit{GALEX} MIS, VST KiDS, VISTA VIKING, \textit{WISE, Herschel-ATLAS}, GMRT and ASKAP providing UV to radio coverage.
GAMA is funded by the STFC (UK), the ARC, the AAO, and the participating institutions.
JBH acknowledges support under an ARC Laureate Fellowship.
JTA and ISK acknowledge the support of SIEF John Stocker Fellowships. 
SMC acknowledges the support of an ARC Fellowship (FT100100457).
LC acknowledges support under the ARC's Discovery Projects funding scheme (DP130100664).
BC is the recipient of an ARC Future Fellowship (FT129199660).
MSO acknowledges funding support from the ARC Super Science Fellowship (ARC FS110200023).

\section*{Appendix A}
Here we summarise properties of some GAMA sub-sample galaxies showing significant $r$-band photometric asymmetries after subtracting a single-\sersic\ profile.
We distinguish between those with and without streaming by bars or spiral arms, ordered as in Figs.~12 and 13 from smallest to largest peak CSC velocity.
As noted below, \diskfit sometimes found no significant kinematical distortion as evidenced by opposite signs for $V_{\rm 2,r}$ and $V_{\rm 2,t}$ regardless of prior bar PA and extent chosen.
For those, Fig.~12 shows the result aligned with the photometric feature to gauge uncertainties.
Where feasible, we classified morphology \citep{Buta07,Kormendy12}, with the caveat that SDSS and VIKING images are too shallow to see outer rings that would anyway be outside the SAMI aperture. 
Fig.~\ref{fig:streamlines} shows the deprojected, face-on, square-root intensity scaled, non-axisymmetric residuals after subtracting the dual-\sersic\ fitted profile.
We also show the best-fit elliptical streamlines and modal content.
The CSC is rarely modified even at the bar radii.

Unless noted, emission-line flux ratios (generally [\ion{S}{ii}]/\ha\ vs. [\ion{O}{iii}]/\hb) indicate insignificant gaseous excitation by shocks, unsurprising given the generally low $m=2$ in-plane velocities from our fits (with uncertainties from spatial averaging over kinematical twists). 
Shocks noted have emission-line ratios consistent with shock velocities of 
$150-300$ \kms and are plotted deprojected as coloured rectangles in Fig.~13.
Radial extents on sky are reported. 
A large span in $m=2$ uncertainties in Fig.~\ref{fig:barfits} generally arose when the kinematic centre and/or bar PA were unconstrained in our fit.
The latter could occur when our fits were averaged by being constrained inappropriately to follow a fixed kinematical PA.

\paragraph*{279818}
A ``flocculent'' Sc with $\log_{\rm 10}(M_{\rm H~I}/M_\odot)=9.35$; only 10 percent H~I profile flux asymmetry.
$m=2$ reached only $\sim5$ \kms\ at 3 arcsec along PA 3\arcdeg, so was consistent with spiral-arm streaming. Stubby arms spiral from the bar tips, with one showing emission-line flux ratios from excitation by high-velocity shocks well above the tiny streaming motions.
\paragraph*{383259}
SB(s) with perpendicular bar-within-bar, a good candidate to study at higher spatial resolution.
Our fit in Fig.~\ref{fig:barfits} found the PA of the inner bar, with streaming amplitudes never exceeding 20 \kms.
$\log_{\rm 10}(M_{\rm H~I}/M_\odot)=10.18$.
\paragraph*{599582}
Single-\sersic\ fit only, because the bar along PA 165\arcdeg\ is embedded in a strong lens and transects a ring, classic SB(r).
A few spaxels show shock excitation.
\paragraph*{319018}
A compact bulge SB(s) has knots aligned along PA 150\arcdeg\ after the second \sersic\ is subtracted, as frequently seen at bar \textit{ansae} \citep[e.g.]{Kormendy12}.
There is shock emission nearby.
$m=2$ is only a few \kms.
\paragraph*{485885}
Non-axisymmetric motions at 2--3\arcsec\ along PA 15\arcdeg\ reached 30 \kms\ but
were perpendicular to the photometric bar.
So, we detected streaming by the strong bisymmeric spiral arms in this SB(s).
\paragraph*{595060}
SB(rs) with weak bar out to 3\arcsec\ along PA 45\arcdeg. $m=2$ declines from 35 \kms as radius increases.
Shock emission only on the nucleus.
Beyond the bar there are residuals of $\sim8$ \kms\ along the spiral arms.
\paragraph*{209807}
If the $m=2$ distortion PA is allowed to vary, the fit PA is indeterminate over 58$\pm$21\arcdeg\ hence produces uncertain velocities with medians that reach 35 \kms. This perhaps SB(rs) shows strong shock excitation at the ends of the bar.
\paragraph*{619095}
Prominent bar in this SB(s) along PA 110\arcdeg\ with bar motions up to 35 \kms\ out to 3\arcsec.
Residuals of $\sim18$ \kms\ are found along the spiral arms.
\paragraph*{376121}
In this SAB(s) the bar emerges from a compact bulge along PA -5\arcdeg. $m=2$ deviations are $\sim60$ \kms\ at the edge of the SAMI PSF then drop abruptly to remain $<30$ \kms by 4\arcsec.
Significant shock excitation is evident out to 5\arcsec.  
\paragraph*{599761}
Bar along PA 30\arcdeg\ out to 3\arcsec\ in this SB(s), maximum deviation 20 \kms.
A few shocked spaxels near the bar.
\paragraph*{599761}
Bar motion detected only at smallest radius barely outside SAMI PSF along PA -149\arcdeg.
\paragraph*{Galaxies with photometric asymmetries but no bar streaming detected:}
\paragraph*{209701}
Prominent bar residual in this SB(r) along PA 20\arcdeg, out to 3\arcsec\ photometrically, but kinematical residuals $<10$\kms. 
\paragraph*{517070}
Compact bulge with bar along PA 115\arcdeg\ that crosses a ring, SB(rs). Shock excitation along one spiral arm but not near the bar.
\paragraph*{79635}
SAB(r) with weak bar out to 4\arcsec\ along PA 90\arcdeg but $m=2$ motions are in the noise.
There are residuals of $\sim10$ \kms\ along the spiral arms.
$\log_{\rm 10}(M_{\rm H~I}/M_\odot)=10.21$, 20 percent profile flux asymmetry.
\paragraph*{508421}
Prominent bar out to 3\arcsec\ along PA 55\arcdeg, surrounded by broken ring, SB(r).
\paragraph*{227572}
SB(s). Dual-\sersic\ fit indicates a bar along PA 170\arcdeg.
\paragraph*{209181}
Weak bar.

\bsp
\label{lastpage}
\end{document}